\documentclass[11pt]{article}
\usepackage{amsmath,amssymb,color,epsfig,cite}
\usepackage{graphicx}
\usepackage{subfigure}
\usepackage{setspace}
\usepackage{braket}

\usepackage{amsfonts}

\newcommand{\hoch}[1]{$\, ^{#1}$}
\allowdisplaybreaks

\makeatletter
\@addtoreset{equation}{section}
\makeatother

\newcommand{\be}{\begin{equation}}
\newcommand{\ee}{\end{equation}}
\newcommand{\bea}{\setlength\arraycolsep{2pt} \begin{eqnarray}}
\newcommand{\eea}{\end{eqnarray}}
\newcommand{\nn}{\nonumber}

\def\0{{\sst{(0)}}}
\def\1{{\sst{(1)}}}
\def\2{{\sst{(2)}}}
\def\3{{\sst{(3)}}}
\def\4{{\sst{(4)}}}
\def\5{{\sst{(5)}}}
\def\6{{\sst{(6)}}}
\def\7{{\sst{(7)}}}
\def\8{{\sst{(8)}}}
\def\sst#1{{\scriptscriptstyle #1}}

\begin{document}

\begin{center}
{\large {\bf Holographic Entanglement Entropy and Van der Waals transitions in Einstein-Maxwell-dilaton theory}}

\vspace{15pt}

{\large Shou-Long Li\hoch{1,2}  and Hao Wei\hoch{1}}

\vspace{15pt}

\hoch{1}{\it School of Physics, Beijing Institute of Technology, Beijing 100081, China }

\vspace{10pt}

\hoch{2}{\it Center for Joint Quantum Studies and Department of Physics,
School of Science, Tianjin University, Tianjin 300350, China }

\vspace{10pt}

\vspace{15pt}

\vspace{40pt}

\underline{ABSTRACT}
\end{center}

According to the gauge/gravity duality, the Van der Waals transition of charged AdS black holes in extended phase space is conjectured to be dual to a renormalization group flow on the space of field theories. So exploring the Van der Waals transition is potentially valuable for studying holographic properties of charged black hole thermodynamics. There are different transition behaviors for charged dilatonic AdS black holes in Einstein-Maxwell-dilaton (EMD) theory with string-inspired potential with different dilaton coupling constants in diverse dimensions. In this work, we find  a special class of charged dilatonic AdS black holes which have the standard Van der Waals transition. We study the extended thermodynamics of the special class of  black holes, which, in the extremal limit, have near-horizon geometry conformal to AdS$_2 \times S^{D-2}$. We find that, for these black holes, both the pressure-volume transition in fixed charge ensemble and the inverse temperature-entropy transition in fixed pressure ensemble have the standard Van der Waals behaviors. We also find the holographic entanglement entropy undergoes the same transition behaviors for the same critical temperature in fixing the thermodynamic pressure ensemble.

\vfill

 sllee\_phys@bit.edu.cn \ \ \ haowei@bit.edu.cn

\thispagestyle{empty}

\pagebreak



\newpage
\tableofcontents
\addtocontents{toc}{\protect\setcounter{tocdepth}{2}}

\section{Introduction}

 According to gauge/gravity duality, also known as anti-de Sitter/conformal field theory (AdS/CFT) correspondence~\cite{Maldacena:1997re, Gubser:1998bc, Witten:1998qj}, there are deep connections between anti-de Sitter (AdS) black holes and strongly coupled large $N$ field theory. The dual CFT of Hawking-Page phase transition~\cite{Hawking:1982dh} of Schwarzchild AdS black hole was interpreted clearly as a confinement/deconfinment phase transition by Witten~\cite{Witten:1998zw}. However, the exact dual field interpretation of Van der Waals transition of Reissner-Nordstr\"om (RN) AdS black hole is still unknown. Actually, the phase transitions of charged AdS black holes have been studied in many literatures, e.g.~\cite{Chamblin:1999tk, Chamblin:1999hg, Cvetic:1999ne, Cvetic:1999rb}. Recently, the idea of treating cosmological constant and its conjugate quantity as thermodynamic pressure and thermodynamic volume in first law of black hole thermodynamics were studied in Refs.~\cite{Caldarelli:1999xj,Kastor:2009wy, Cvetic:2010jb} and then the phase transition in extended phase space was considered in Ref.~\cite{Kubiznak:2012wp}. There appear several new interesting transition behaviors, such as $\lambda$-line transition~\cite{Hennigar:2016xwd}, reentrant phase transitions~\cite{Gunasekaran:2012dq, Altamirano:2013ane, Frassino:2014pha, Zou:2016sab, Dehyadegari:2017hvd}, triple points~\cite{ Altamirano:2013uqa, Frassino:2014pha, Wei:2014hba}, special isolated critical point~\cite{Dolan:2014vba}, transition at negative pressure~\cite{Li:2018rpk} and so on~\cite{Hendi:2017fxp, Caceres:2015vsa} in extended phase space. We refer to e.g.~\cite{Altamirano:2014tva, Kubiznak:2016qmn} and references therein for more details of this subject.

Recently, a conjecture~\cite{Johnson:2014yja} was proposed that the Van der Waals transition in extended phase space is related to renormalization group (RG) flow which is an holographic explanation of extended thermodynamics of black hole. Some other related efforts, e.g.~\cite{Caceres:2015vsa, Johnson:2015ekr, Johnson:2015fva, Johnson:2016pfa, Bhamidipati:2016gel, Chakraborty:2016ssb, Mo:2017nhw,  Hennigar:2017apu, Liu:2017baz, Johnson:2017ood,  Mo:2017nes, Hendi:2017bys, Wei:2017vqs, Chakraborty:2017weq, Zhang:2018vqs, Rosso:2018acz, Johnson:2018amj, Zhang:2018hms}, have been made in such interpretation. So as the first step to study the holographic properties of thermodynamics of charged AdS black holes, exploring the Van der Waals behaviors in corresponding gravities is worth to do. Actually, the standard Van der Waals behavior does not always exist in charged AdS black hole systems. For examples, according to recent works~\cite{Caceres:2015vsa, Li:2018rpk}, there are different phase transition behaviors in charged dilatonic AdS black holes in a class of Einstein-Maxwell-dilaton (EMD) theory, which can be embedded in gauged supergravites, for different dilaton coupling constant $a$, also known as $N$, in diverse dimensions.

The extremal RN black holes in supergravity can be viewed as one of the bound states of the basic $U(1)$ building blocks with zero binding energy~\cite{Duff:1994jr, Rahmfeld:1995fm}. On the other hand, while EM theories can be embedded in string and M-theory in four and five dimensions only, charged dilatonic AdS black holes in gauged supergravities~\cite{Behrndt:1998jd, Duff:1999gh, Cvetic:1999xp, Cvetic:1999un} can be embedded in higher dimensions. The considered EMD theory with string-inspired potential, as a concrete gravitational theory, can be generalized to arbitrary dimensions~\cite{Lu:2013eoa}.

From Refs.~\cite{Caceres:2015vsa, Li:2018rpk}, we find that only the four-dimensional $N=3$ and five-dimensional $N=2$ cases have the traditional Van der Waals behavior which is the same as the RN-AdS black hole in the frame of gauge supergravity. It is natural and valuable to ask whether the Van der Waals behaviors of $D=4, N=3$ and $D=5, N=2$ cases are coincidences. Is there some deep reason for the existence of Van der Waals behaviors in four and five dimensions? Can we observe the same behaviors in higher than five dimensions in the EMD theory with supergravity properties?
We find charged dilatonic AdS black holes in $D=4, N=3$ and $D=5, N=2$ cases share a common point that having near-horizon geometry conformal to AdS$_2 \times S^{D-2}$ in the extreme limit~\cite{Lu:2013eoa, Li:2018omr}, while extremal RN-AdS black holes have near-horizon geometry AdS$_2 \times S^{D-2}$~\cite{ Kunduri:2013ana}. So we conjecture the similar near-horizon geometry may be related to the same behavior as the phase transition. In order to demonstrate our conjecture, we generalize previous transitions from gauged supergravity to EMD theory with supregravity properties in general dimensions. For this special class of black holes with near-horizon geometry conformal to AdS$_2 \times S^{D-2}$, the phase transitions in fixing both the charge and thermodynamic pressure ensembles are indeed standard Van der Waals behaviors.

On the other hand, holographic entanglement entropy (HEE) which enjoys much attention since proposed by Ryu and Takayanagi~\cite{Ryu:2006bv, Rangamani:2016dms} undergoes also standard Van der Waals behaviors in RN AdS black holes instead of Bekenstein-Hawking entropy~\cite{Johnson:2013dka}. The HEE transitions were also studied in many other modified gravities, i.e.~\cite{Nguyen:2015wfa, Caceres:2016xjz, Couch:2016exn, Liu:2017jbm, Zeng:2017zlm, Zeng:2016fsb, Sun:2016til, Hendi:2016uni, Zeng:2016aly, Mo:2016cmi, Zeng:2016sei, Kundu:2016dyk, Zeng:2015wtt, Li:2017xiv, Li:2017gyc, Zeng:2015tfj, Dudal:2016joz, Sinamuli:2017rhp}. And the Van der Waals transition behavior in charged dilatonic AdS black holes in four-dimensional gauged supergravity also exists in $N=3$ case only~\cite{Caceres:2015vsa}. So it is natural to check whether Van der Waals transition behavior exists in the special class charged dilatonic AdS black holes with conformal AdS$_2 \times S^{D-2}$ near-horizon geometry in general dimensions.

The paper is organized as follows. In section~\ref{sec2}, we review the thermodynamics of dilatonic charged AdS black holes in EMD theory with string-inspired potential and study the near-horizon geometry properties of the special class black holes. In section~\ref{sec3}, we study the extended thermodynamics of the special class black holes, especially thermodynamic pressure versus volume transition in fixing the charge ensemble and the inverse temperature versus Bekenstein-Hawking entropy in fixing pressure ensemble. We find the transitions have standard Van der Waals behaviors in both cases. In section~\ref{sec4}, we review the HEE and calculate it numerically. Then we study the inverse temperature versus the HEE transition in fixing the pressure ensemble. We find the standard Van der Waals behaviors still exist. We conclude in section~\ref{sec4}.

\section{Charged dilatonic AdS black holes with conformal AdS$_{2} \times S^{D-2}$ near-horizon geometry } \label{sec2}

The Lagrangian of the general EMD theory consisting of gravity, a single Maxwell field $A$, and a dilaton field $\phi$ in $D\ge 4$ dimensions is given by
\be
e^{-1} {\cal L}= R-\frac12(\partial\phi)^2 -\frac14 e^{a\phi} F^2 -V(\phi)  \,, \label{lagrangian}
\ee
where $e=\sqrt{-g}$, $F = d A$, $V$ is scalar potential inspired by gauged supergravities.  For later purposes, it is convenient to reparametrize the dilaton coupling constant $a$ by~\cite{Lu:2013eoa}
\be
a = \sqrt{\frac{4}{N}-\frac{2 (D-3)}{D-2} } \,.\label{constraint1}
\ee
The reality condition of $a$ requires that the constant $N$ must satisfy
\be
0<N\leq N^{RN} \,, \quad \textup{with} \quad N^{RN}= \frac{2 (D-2)}{D-3} \,.  \label{reality}
\ee
The charged dilatonic AdS black holes with $a$ can be viewed as dilatonic AdS black hole with $N$ equal charges~\cite{Duff:1994jr, Rahmfeld:1995fm}. So the value of $N$ should be positive integers required by supergravity.  However, as a concrete gravitational theory, any value of $N$ satisfying Eq.~(\ref{reality}) is allowed. When $N=N^{RN}$, i.e., $a=0$, the dilaton decouples and the theory reduces to EM theory. Although the Lagrangian can be made real by letting $\phi\rightarrow i \phi$ when $N>N^{RN}$, we shall not consider such a situation at all. The potential can be expressed in terms of a super potential $W$~\cite{Lu:2013eoa},
\be
V= \left( \frac{dW}{d\phi} \right)^2 -\frac{D-1}{2(D-2)}W^2 \,,\quad \textup{with} \quad W= \frac{1}{\sqrt{2}} N (D-3) g \left(e^{-\frac12 a \phi} -\frac{a}{\tilde{a}} e^{-\frac12 \tilde{a} \phi} \right) \,, \label{potential}
\ee
where $\tilde{a} a = -2(D-3)/(D-2)$ and $g$ is the gauge coupling constant~(there should be no confusion between the gauge coupling constant and the determinant of the metric).

The static AdS black hole solutions for the Lagrangian~(\ref{lagrangian}) with scalar potential~(\ref{potential}) and constraints~(\ref{constraint1}) are given by~\cite{Lu:2013eoa}
\begin{align}
ds^2 &= -h^{-\frac{D-3}{D-2} N} f dt^2 + h^{\frac{N}{D-2}} \Big(\frac{dr^2}{f}+r^2 d\Omega_{D-2}^2\Big) \,,\label{bhsol} \\
A &=\sqrt{\frac{N(m+q)}{q}} h^{-1} dt \,,\quad \phi = \frac12 N a \log h \,,\\
f &= 1-\frac{m}{r^{D-3}}+ g^2 r^2 h^{N} \,, \quad h = 1+\frac{q}{r^{D-3}} \,,
\end{align}
where parameters $m$ and $q$ characterize mass and electric charge, and $d\Omega_{D-2}^2$ represents the unit $(D-2)$-sphere, $(D-2)$-torus or hyperbolic $(D-2)$-space. The topological black holes
can be easily obtained by some appropriate scaling. In our work, we only consider spherical black holes for simplification.

The event horizon of the black hole is determined by the largest (real) root of $f(r_0) = 0$. The thermodynamic quantities are given by~\cite{Lu:2013eoa, Feng:2017wvc}
\begin{align}
M &= \frac{\pi^{\frac{D-3}{2}}}{8 \, \Gamma(\frac{D-1}{2})} \left( (D-2) m + (D-3) N q \right) \,, \label{mass} \\
T &= \frac{f'}{4 \pi  h^{\frac{N}{2}}}  \,,  \qquad S = \frac{1}{4} {\cal A} = \frac{\pi^{\frac{D-1}{2}} r_0^{D-2}}{2 \, \Gamma(\frac{D-1}{2})} h^{\frac{N}{2}} \,,  \label{temp} \\
Q &= \frac{(D-3) \pi^{\frac{D-3}{2}}}{8 \, \Gamma(\frac{D-1}{2})} \sqrt{N q (m+q)} \,, \qquad \Phi = \sqrt{\frac{N (m+q)}{q}} \left(1-\frac{1}{h} \right) \,, \label{charge}
\end{align}
where $\Gamma$ indicates the gamma function, and $(M, T, S, Q, \Phi)$ denote mass, Hawking temperature, entropy, electric charge and electric potential respectively.

\begin{figure}[h]
\centerline{
\ \ \ \ \ \includegraphics[width=0.3\linewidth]{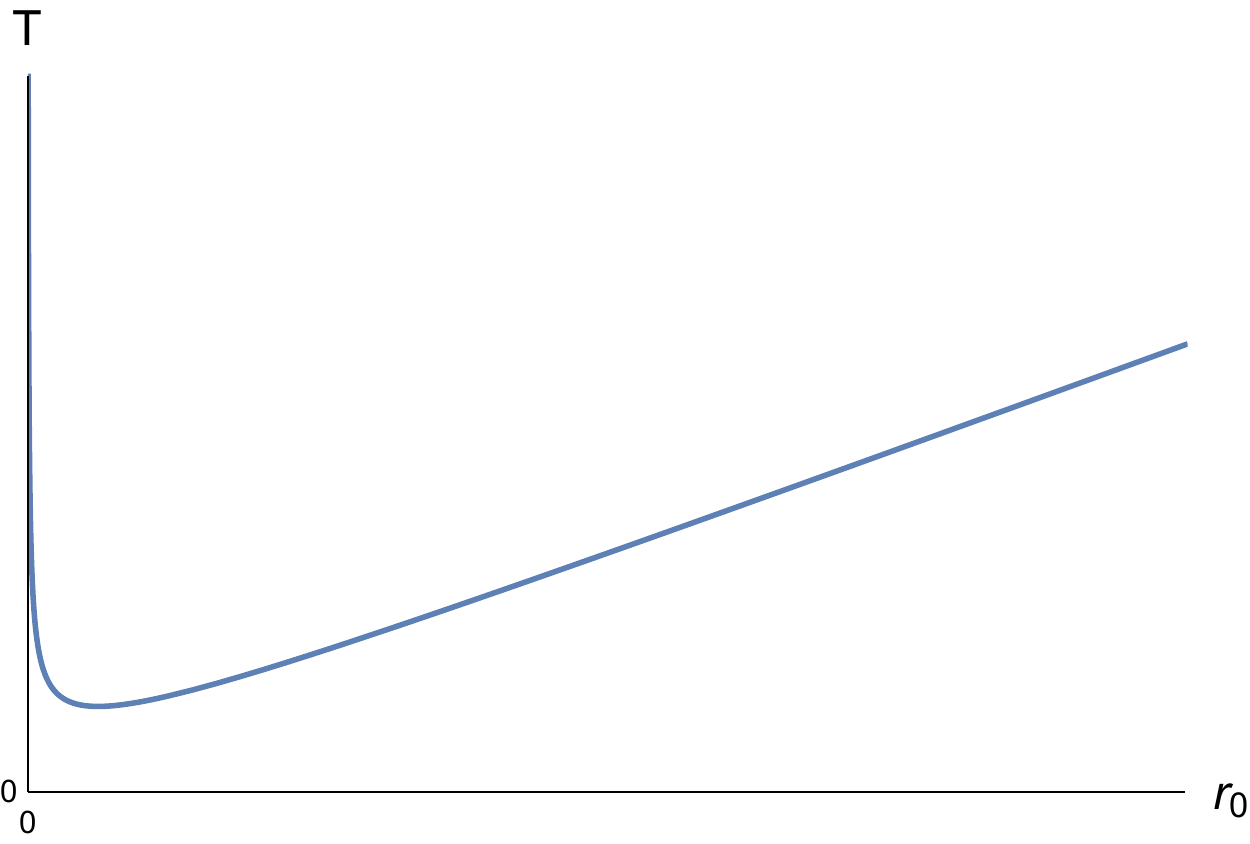}\ \ \ \ \ \includegraphics[width=0.3\linewidth]{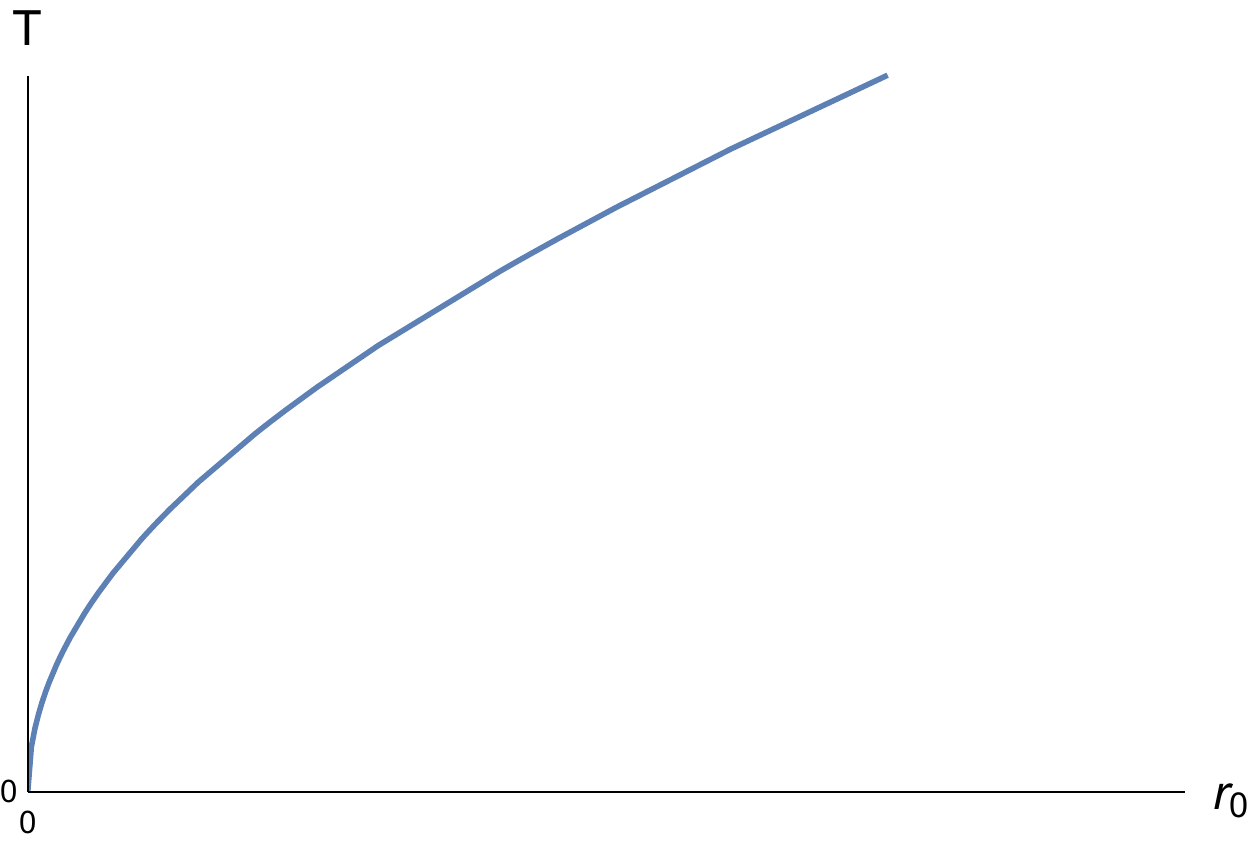}\ \ \ \
\includegraphics[width=0.3\linewidth]{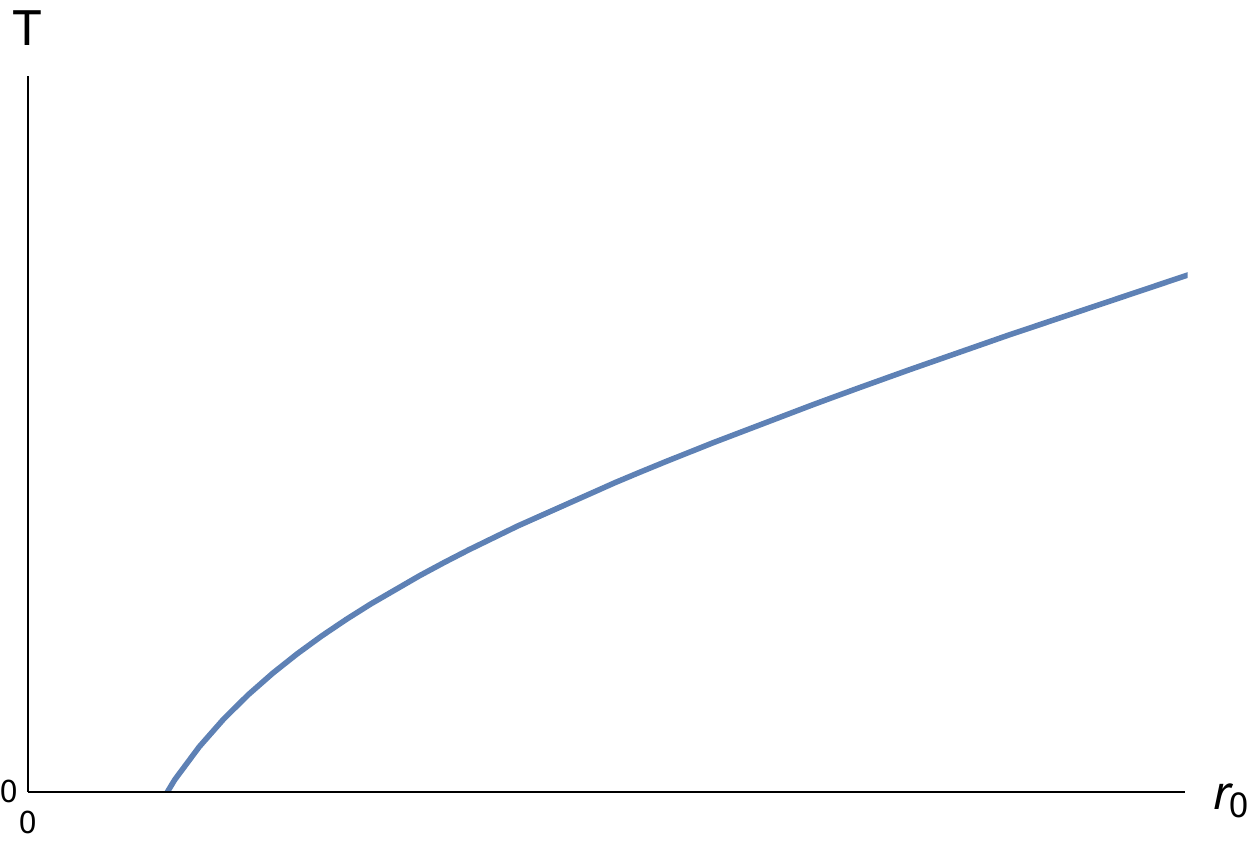}}
\caption{Qualitative relations between $T$ and $r_0$ for different values of $N$ by fixing the charge parameter $q$ and gauge coupling constant $g$. Left: $0 <N <N^\star$; middle: $N= N^\star$; right: $N^\star <N \le N^{RN}$.  }
\label{Tr0nvalue}
\end{figure}

The extremal black holes can be obtained by requiring $f(r_0) = f^\prime (r_0) =0$, but with $f^{\prime\prime} (r_0) \ne 0$. The extreme limit is also called zero temperature limit. The extremal black holes in EMD theory with two Maxwell field~\cite{Lu:2013eoa} always exist for any dilaton coupling constant because the theory reduces to EM theory when the two Maxwell field are equal. However, the existence of extremal black hole in EMD theory with single Maxwell field~(\ref{lagrangian}) depends on the value of $N$. For general $N$,  $f^\prime (r_0)$ becomes
\be
f^\prime (r_0) = (D-1 -N (D-3)) g^2 r_0 (1+\frac{q}{r_0^{D-3}})^N  + \frac{D-3}{r_0} +N(D-3) g^2 r_0 (1+\frac{q}{r_0^{D-3}})^{N-1}   \,.
\ee
According to Eq.~(\ref{temp}), the temperature is given by
\be
\begin{split}
T &= \frac{D-3}{4 \pi } \Big[ \Big(\frac{D-1}{D-3} -N \Big) (q+ r_0^{D-3 } )^{\frac{N}{2}}  g^2 r_0^{\frac{2-(D-3) N}{2}}  + (q+ r_0^{D-3 } )^{-\frac{N}{2}} r_0^{-\frac{2-(D-3) N}{2}}  \\
&\quad +N g^2 (q + r_0^{D-3})^{\frac{N-2}{2}} r_0^{\frac{D-3}{2} (N^{RN}-N)}  \Big] \,.
\end{split}
\ee
It is easy to see that the last two terms in the bracket in the above equation are always greater than zero. There is a special value
\be
 N^\star=\frac{D-1}{D-3} \,, \quad \textup{with} \quad a = \sqrt{\frac{2}{(D-1)(D-2)}}\, (D-3) \,. \label{Nstar}
\ee
For $N <N^\star$, $T $ is always greater than 0. For $N^\star <N \le N^{RN}$, there must exist extremal black holes for proper parameters. If we choose $N = N^\star $, there is also a extreme limit as $r_0 \rightarrow 0$. The special dilaton coupling constant can be viewed as a bound threshold for the existence of extremal black hole. We plot the relations between $T$ and $r_0$ for both cases in Fig.~\ref{Tr0nvalue}.
 In this case, the lowest order of $r$, $1/r^{D-3}$, in $f$ cancels for $m = g^2 q^{N^\star}$ and $f$ becomes a nonvanishing constant as $r \rightarrow 0$. The $r_0 =0$ is then a null horizon where the horizon and curvature singularity coincide. In this case, near the null horizon, $r\rightarrow 0$, the solution~(\ref{bhsol}) becomes
\bea
ds^2 &=& -f_0 (\frac{q}{r^{D-3}})^{-\frac{D-1}{D-2}} dt^2 +f_0^{-1} (\frac{q}{r^{D-3}})^{\frac{D-1}{(D-2)(D-3)}} dr^2 +(\frac{q}{r^{D-3}})^{\frac{D-1}{(D-2)(D-3)}} r^2 d\Omega_{D-2}^2  \,\nn\\
&=& r^{\frac{D-3}{D-2}} (-c_1 r^{D-3} dt^2 +c_2 r^{-2} dr^2 +c_3 d\Omega_{D-2}^2 )  \,\nn\\
&=& \tilde{r}^{-\frac{2}{D-2}} (\frac{- d\tilde{t}^2 + d\tilde{r}^2}{\tilde{r}^2} + d\tilde{\Omega}_{D-2}^2 ) \,,
\eea
where
\be
f_0 = \frac{D-1}{D-3} g^2 q^{\frac{2}{D-3}} \,, \quad c_1 = f_0\ q^{-\frac{D-1}{D-2}} \,, \quad c_2 = f_0^{-1} q^{\frac{D-1}{(D-2)(D-3)}} \,, \quad c_3 = q^{\frac{D-1}{(D-2)(D-3)}} \,.
\ee
The near-horizon geometry becomes conformal AdS$_{2}\times S^{D-2}$. To be specific, the metric $\tilde{r}^{\frac{2}{D-2}} ds^2$ becomes AdS$_{2}\times S^{D-2}$ around $\tilde{r} =0$. So charged dilatonic AdS black holes with $N= N^\star$ have the similar near-horizon structure with the RN-AdS black holes with $N =N^{RN}$ in EMD theory. In the range of $N^\star <N \le N^{RN}$, the black hole must have the similar behaviors. So we can say the Van der Waals behaviors must exist in the range $N^\star \le N \le N^{RN}$ at least in zero temperature limit. However, at finite temperature, the phase transition behaviors are still unclear. We will study the behaviors in the next section.

\section{Extended thermodynamics }  \label{sec3}

In this section we study the extended thermodynamics of charged dilatonic AdS black holes  in EMD theory with $N = N^\star$. The gauge coupling constant (cosmological constant) can be interpreted as thermodynamical pressure ${\cal P}$ in the extended phase space,
\be
{\cal P} = -\frac{1}{8 \pi} \Lambda = \frac{(D-1)(D-2)}{ 16 \pi} g^2  \,. \label{pressure}
\ee
The corresponding thermodynamic volume is given by
\be
\begin{split}
{\cal V}
= \frac{2 \pi^{\frac{D-1}{2}} \left(q + R\right)^{N-1} }{(D-1)\, \Gamma(\frac{D-1}{2})} \left(  R + \Big(1- \frac{N}{N^{RN}} \Big) q \right) R^{N^\star -N} \,,
\end{split} \label{vol1}
\ee
where $R =r_0^{D-3}$. The thermodynamic volume satisfies the ``reverse isoperimetric inequality'' conjecture~\cite{Cvetic:2010jb}. Although the conjecture is not proven, it has been checked in most black holes and follows from the null-energy condition~\cite{Feng:2017wvc}. In order to study the ${\cal P - V}$ phase transitions, it is convenient to introduce a special volume $v$ by analyzing the dimensional scaling. By assuming the shape of black hole is regular sphere, the special volume can be viewed as the effective radius of black hole,
\be
\frac{2 \pi^{(D-1)/2}}{(D-1) \Gamma(\frac{D-1}{2})} v^{D-1 } = {\cal V} \quad  \Rightarrow \quad   v = \Big(\frac{(D-1) \Gamma(\frac{D-1}{2}) {\cal V}}{2 \pi^{(D-1)/2} } \Big)^{\frac{1}{D-1}}  \,.
\ee
Together with other thermodynamical quantities~(\ref{mass})-(\ref{charge}), we have the following first law and Smarr relation
\bea
d M &=& T dS +\Phi dQ +{\cal V} d{\cal P}  \,,\\
M &=& \frac{D-2}{D-3} T S + \Phi Q + \frac{2}{D-3} {\cal V} d{\cal P} \,.
\eea

We present the values of the special dilaton coupling constant in diverse dimensions in Table~ \ref{table:Nvalue}.
 \begin{table}[h]
 \centering
\begin{tabular}{|c|c|c|c|c|c|c|c|c|c|c|}
   \hline  $D$  & 4 & 5 & 6 & 7 & 8 & 9 & 10 & 11 & $\cdots$ & $\infty$  \\
   \hline $N^\star $ & 3 & 2 & 5/3 & 3/2 & 7/5 & 4/3 & 9/7 & 5/4 & $\cdots$ & 1   \\
   \hline $a $ & $1/\sqrt{3}$ & $\sqrt{2/3}$ & $3/\sqrt{10}$ &$ 4/\sqrt{15} $&$ 5/\sqrt{21} $& $3/\sqrt{7} $& $7/6 $&$ 8/(3\sqrt{5}) $& $\cdots$ & $\sqrt{2}  $ \\
  \hline
\end{tabular}
   \caption{The values of $N^\star $ and corresponding $a$ in diverse dimensions.}   \label{table:Nvalue}
 \end{table}
The four- and five- dimensional cases coincide with gauged supergravities. The theory with $N=1$ case in infinite $D$ dimension (large $D$ limit) is coincident with gauged Kaluza-Klein theory. The ${\cal P} -v$ phase transitions of RN-AdS black holes in the extended phase space undergo the Van der Waals behaviors~\cite{Kubiznak:2012wp}. The similar transition behaviors of charged dilatonic AdS black holes in gauged supergravities have also been studied in Refs.~\cite{Caceres:2015vsa,Li:2018rpk}. There are only two cases $D=4, N =3$ and $D=5, N=2$ that have existing standard Van der Waals behavior. Both cases are subset of  $N = N^\star$  and share the same near-horizon geometry. So we study the ${\cal P }-v$ phase transition in EMD theory with supergravity property with special dilaton coupling constant $N = N^\star$ in diverse dimensions. On the other hand, inverse temperature versus Bekenstein-Hawking entropy and HEE transitions of RN-AdS black holes~\cite{Johnson:2013dka} and STU black holes with three equal charges~\cite{Caceres:2015vsa} also have the Van der Waals behaviors. So we also want to study whether the similar behaviors exist in EMD theory with the special dilaton coupling constant $N = N^\star$. In this section, we study the ${\cal P } -v$ and $T^{-1} -S$ phase transitions of the charged dilatonic AdS black holes in extended phase space. We fix the electric charge $Q$ and thermodynamical pressure $\cal P$ respectively, and do not treat them as thermodynamical variables in each case. The first law reduces to
\be
d M  = T dS +{\cal V} d{\cal P} \,,\quad \textup{and} \quad  d M  = T dS +\Phi d Q \,,
\ee
for each case.

\subsection{${\cal P}-v$ criticality }  \label{sec3.1}

\begin{figure}[h]
\centerline{
\ \ \ \ \ \includegraphics[width=0.45\linewidth]{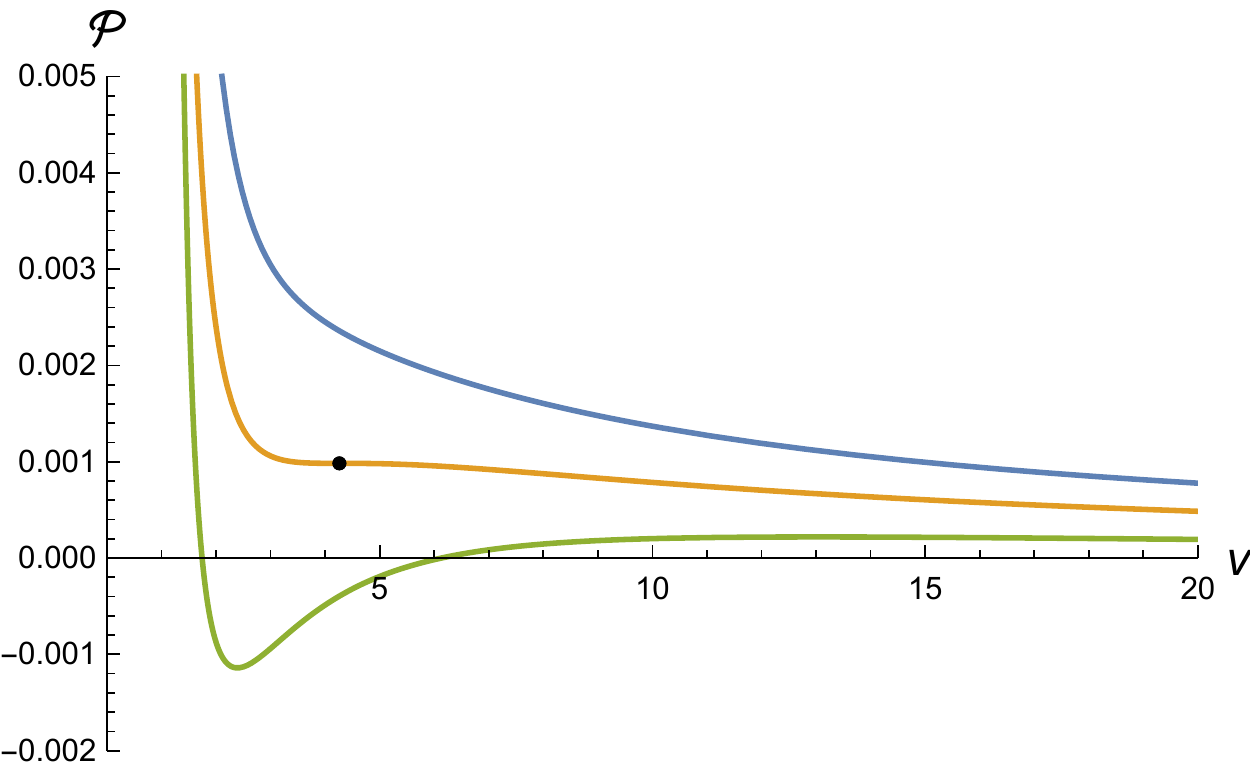}\ \ \ \
\includegraphics[width=0.45\linewidth]{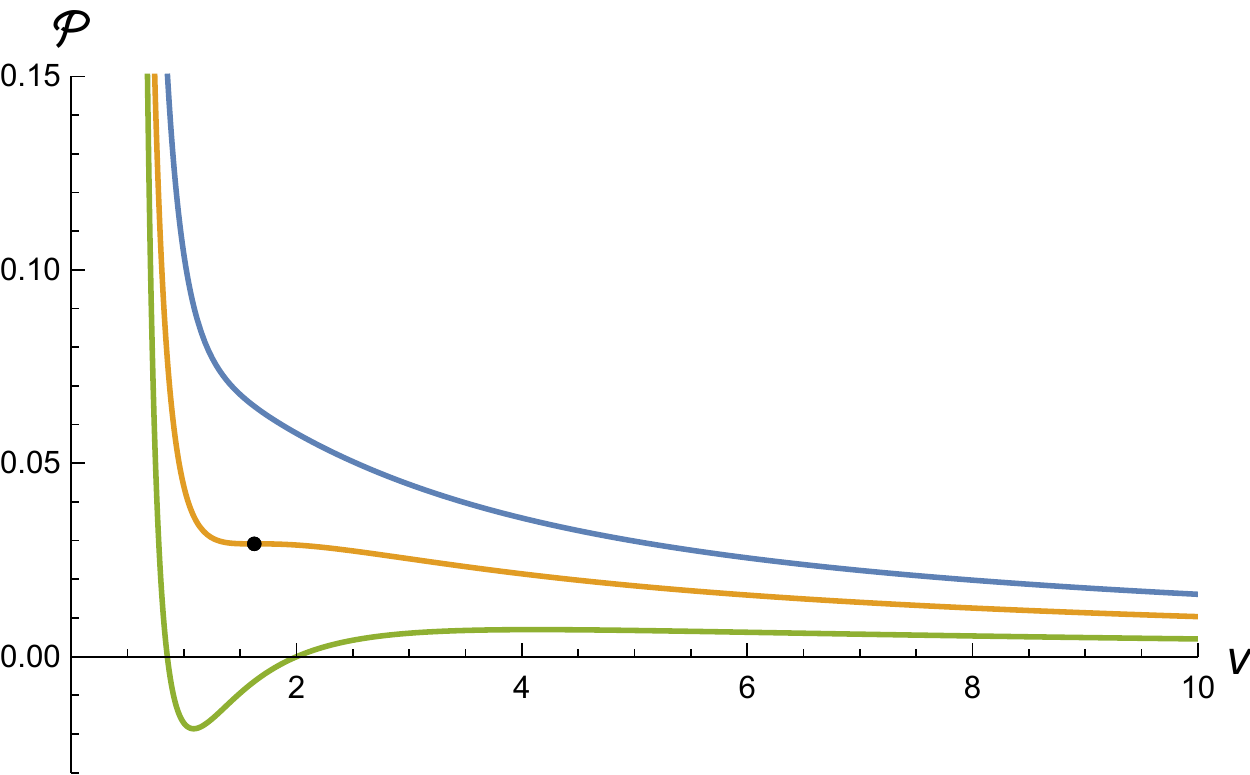}}
\centerline{
\ \ \ \ \ \includegraphics[width=0.45\linewidth]{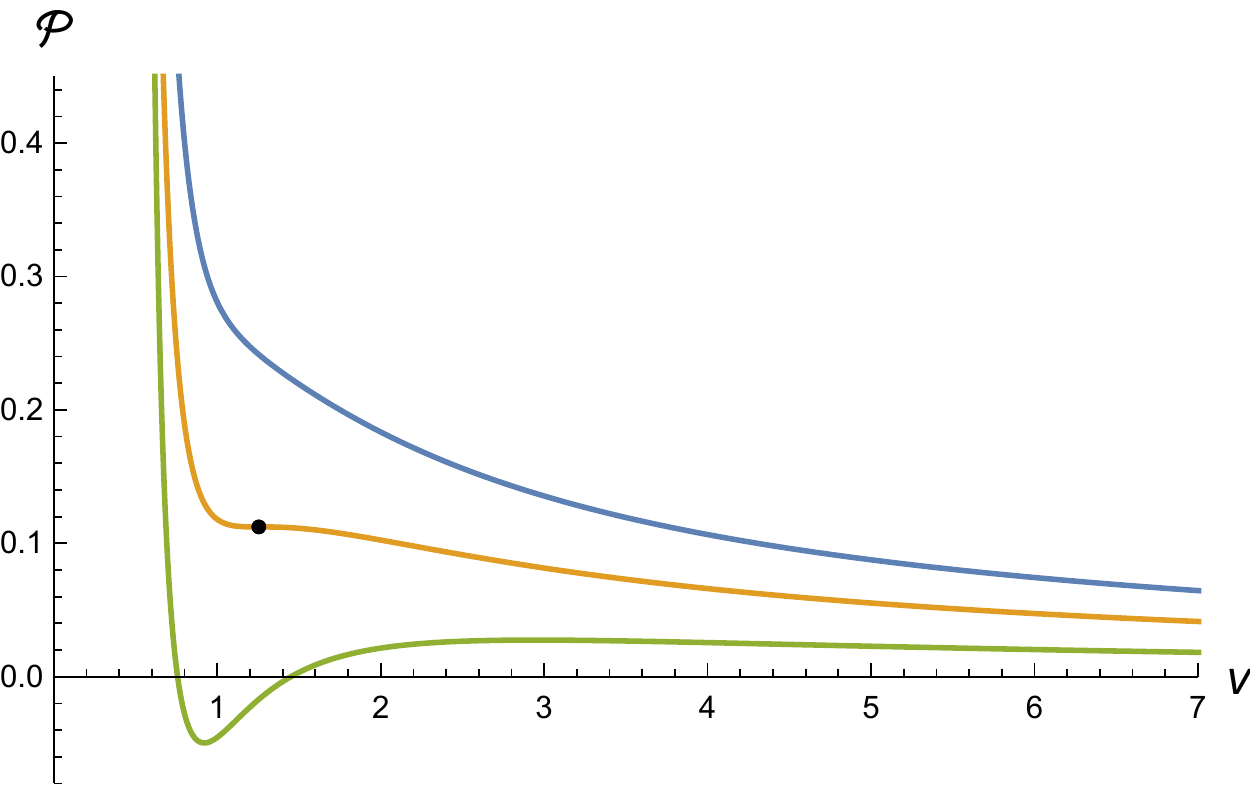}\ \ \ \
\includegraphics[width=0.45\linewidth]{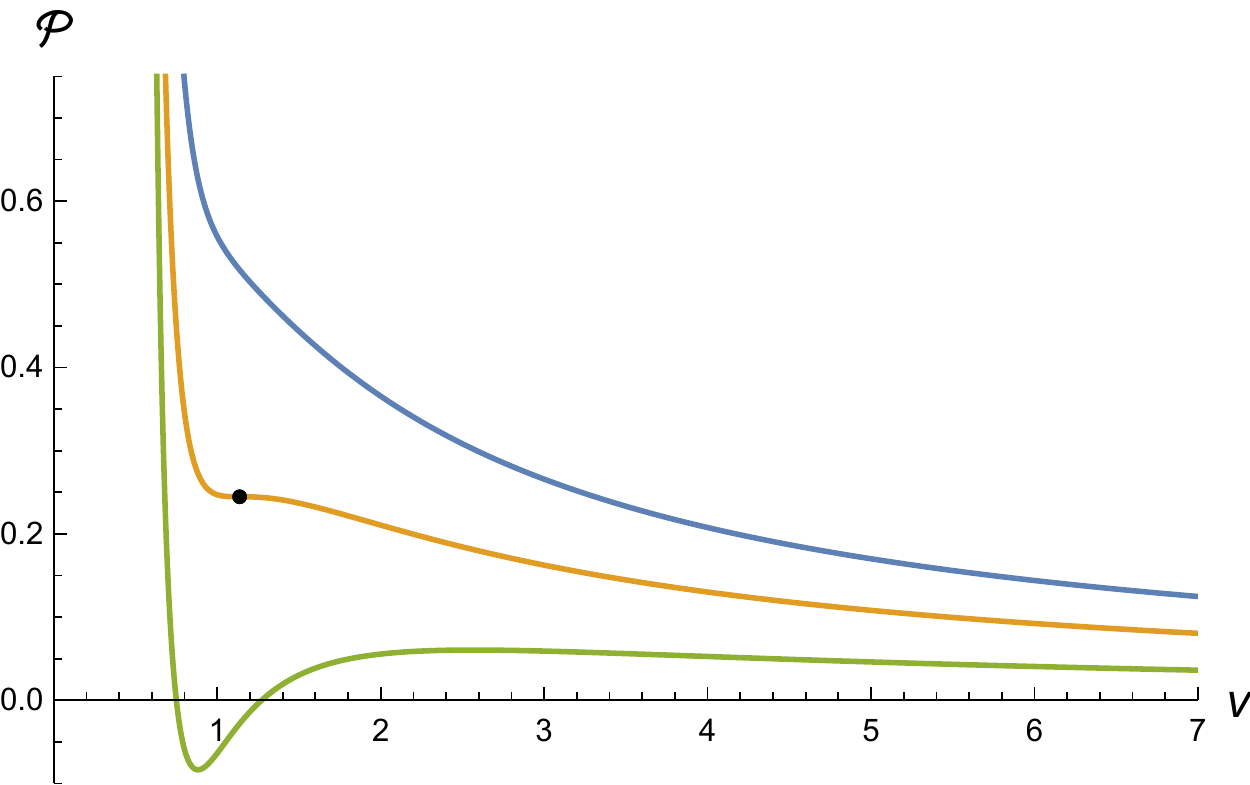}}
\caption{Isotherms in ${\cal P } - v$ diagrams of charged dilatonic AdS black hole in diverse dimensions with $N=N^\star$. We set $Q=1$. Top left: $D=4$; top right: $D=5$; down left: $D=6$; down right: $D=7$.  The blue, orange and green lines represent isotherms with $T = 1.5\ T_c, T_c,$ and $ 0.5\ T_c$ from top to bottom. The black points represent critical points.}
\label{pv56710}
\end{figure}

\begin{figure}[h]
\centerline{
\ \ \ \ \ \includegraphics[width=0.45\linewidth]{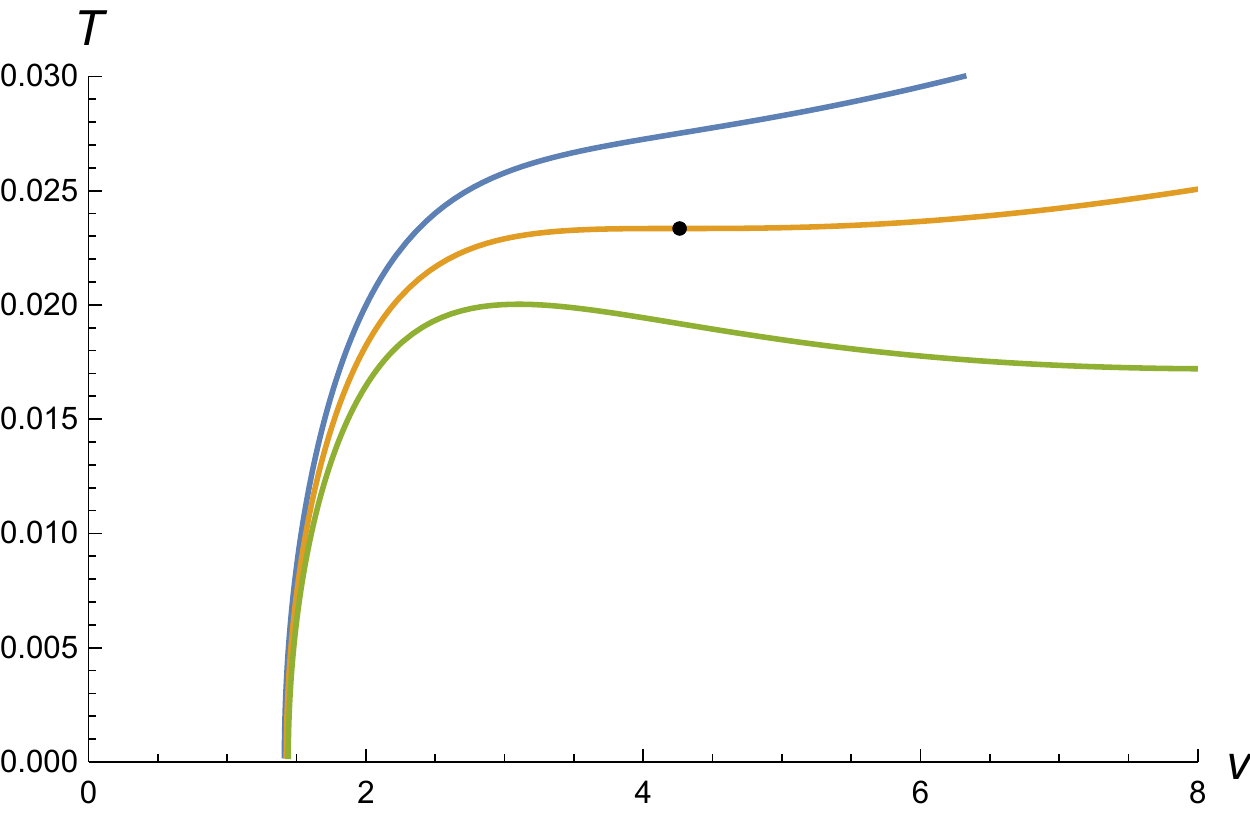}\ \ \ \
\includegraphics[width=0.45\linewidth]{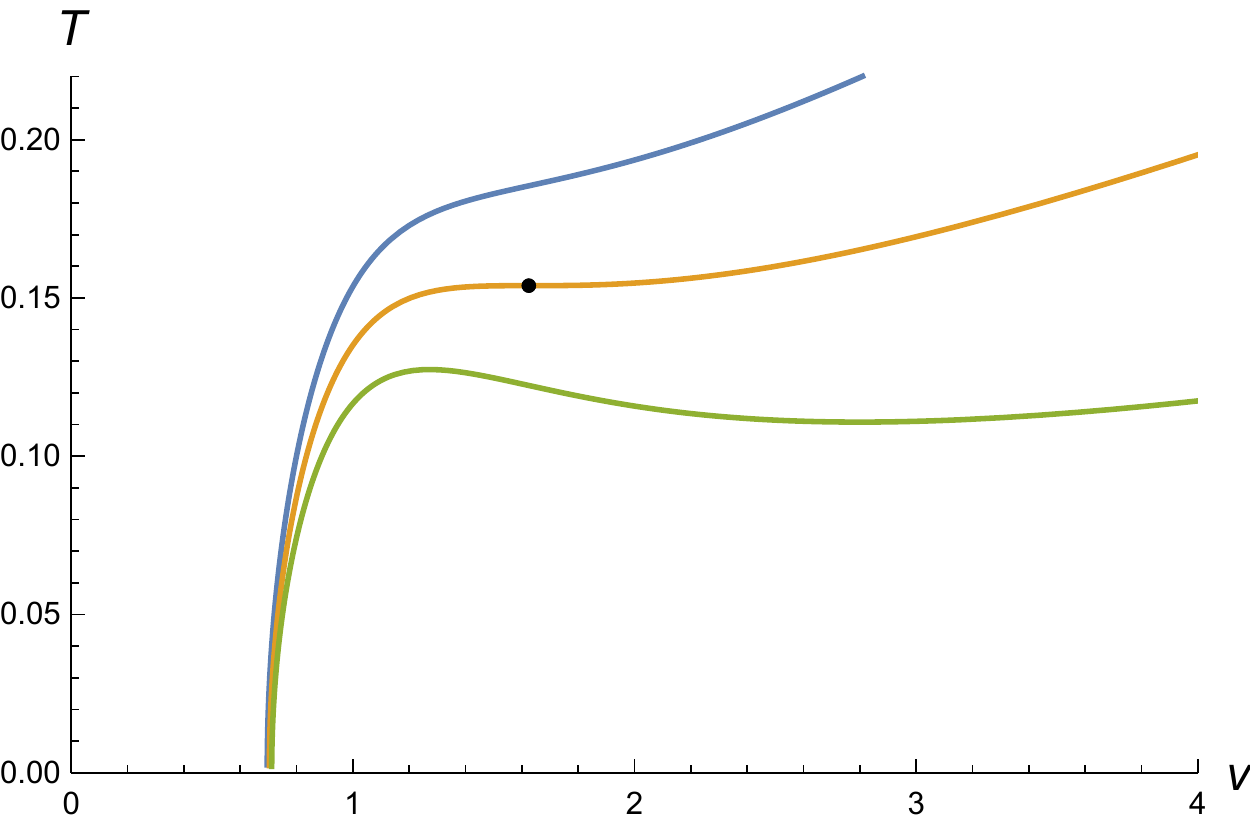}}
\centerline{
\ \ \ \ \ \includegraphics[width=0.45\linewidth]{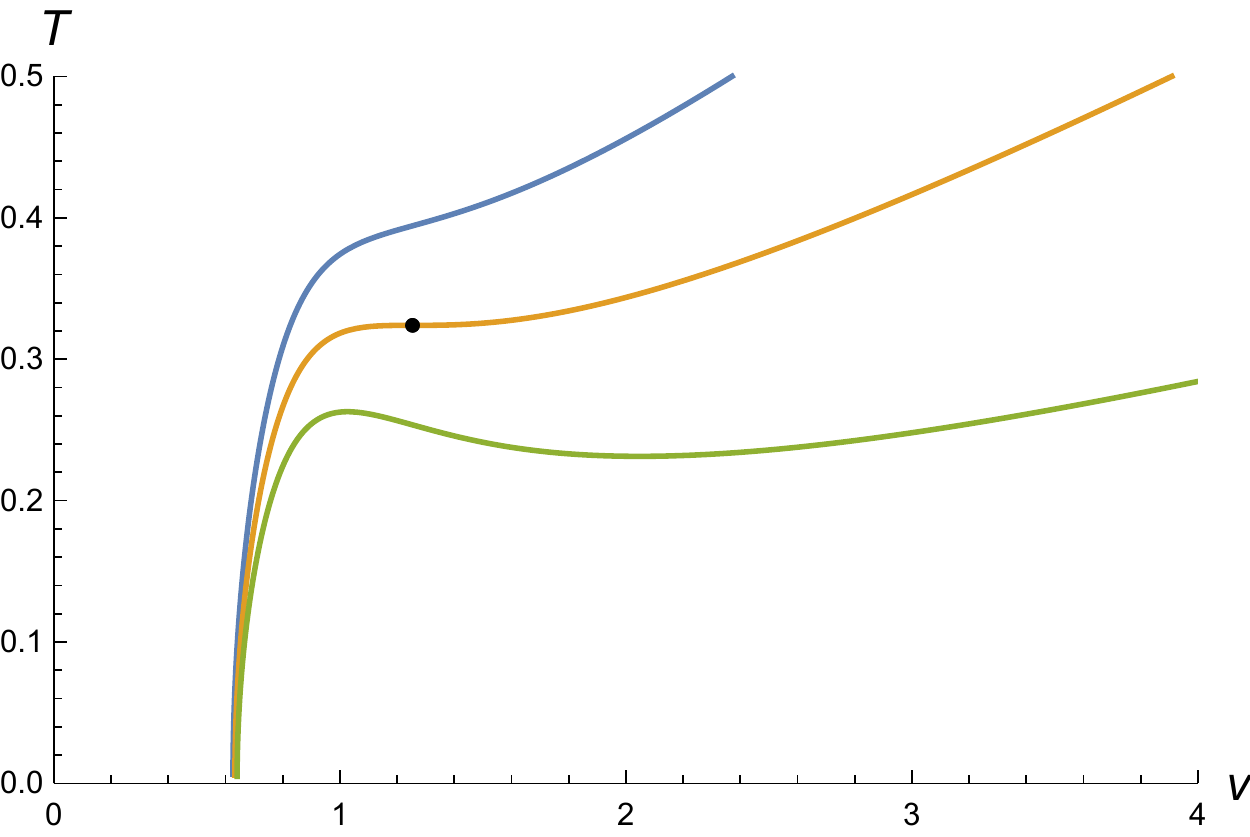}\ \ \ \
\includegraphics[width=0.45\linewidth]{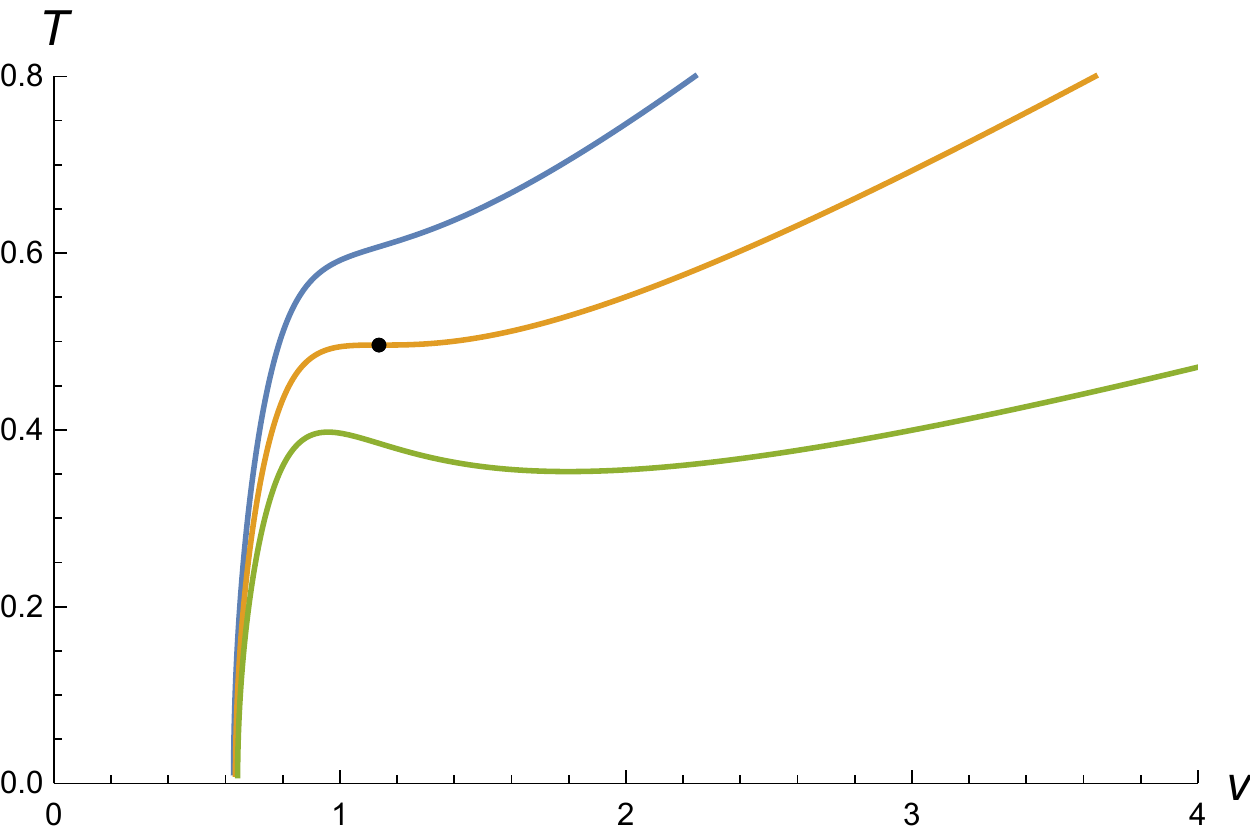}}
\caption{ Isobars in $T - v$ diagrams of charged dilatonic AdS black hole in diverse dimensions with $N=N^\star$. We set $Q=1$. Top left: $D=4$; top right: $D=5$; down left: $D=6$; down right: $D=7$.  The blue, orange and green lines represent isobars with ${\cal P} = 1.5\ {\cal P}_c, {\cal P}_c,$ and $ 0.5\ {\cal P}_c$ from top to bottom. The black points represent critical points.}
\label{vt56710}
\end{figure}

 Firstly, we study the ${\cal P} - v$ transition by treating charge $Q$ as constant. The temperature, thermodynamical pressure and special volume can be written as
\begin{align}
T &= -\frac{R^{\frac{1}{2} }}{2\pi}  (q+R)^{-\frac{N^\star}{2}} +\frac{16 Q^2 \Gamma^2[\frac{D-1}{2}]}{(D-3) \, q\, \pi^{D-2}}\, R^{\frac{1}{2} } {(q +R )^{-\frac{3 D - 7}{2(D-3)}}}   \,,  \label{temp2} \\
P &= \frac{N^{RN}}{32 \pi } \left( \frac{64 Q^2 \Gamma^2[\frac{D-1}{2}]}{q \pi^{D-3} } -(D-1)(D-3)(q+R) \right) \left(q + R\right)^{-N^\star}  \,, \label{pressure2} \\
v &= (q+R)^{\frac{2}{(D-1)(D-3)}} \left( \frac{q}{N^{RN}} +R \right)^{\frac{1}{D-1}} \,.  \label{rho2}
\end{align}
It is hard to obtain the exact EoS because the temperature,  thermodynamical pressure and volume are coupled with each other. However, we can study the phase transition by numeric method. To be specific, we set $Q=1$. The ${\cal P} - v$ diagram is given by Fig.~\ref{pv56710}.
We illustrate only $D=4, 5, 6, 7$-dimensional cases to show the transition behaviors. The ${\cal P} -v$ criticality in four- and five- dimensional cases have been analyzed in Refs.~\cite{Caceres:2015vsa,Li:2018rpk}. However, as the special cases in our analysis, we present them here for comparison. We also explore the transition in higher dimensions and find the behaviors are the same. From the diagram, we find the ${\cal P} -v$ phase transitions in diverse dimensions undergo standard Van der Waals behaviors. As the volume decreases to zero, the pressure increase to infinity. As the volume goes to infinity, the pressure goes to zero. The corresponding critical point can be obtained numerically by
 \be
 \frac{\partial {\cal P}}{\partial v} \Big|_{T_c} = 0 \,, \quad  \frac{\partial^2 {\cal P}}{\partial v^2} \Big|_{T_c}  = 0 \,. \label{cp1}
 \ee
The critical points are given in Table~\ref{table:cp1}.
 \begin{table}[h]
 \centering
\begin{tabular}{|c|c|c|c|c|}
   \hline  $D$  & $T_c$ & $v_c$ & ${\cal P}_c$ & $\frac{{\cal P }_{c} v_{c}}{T_{c}}$ \\
   \hline 4 & $0.0233 \ Q^{-1}$ & $ 4.26 \ Q$ &  $0.000983 \ Q^{-2}$ & 0.179    \\
   \hline 5 & $0.154 \ Q^{-1/2}$ & $ 1.62 \ Q^{1/2}$ & $0.0292 \ Q^{-1}$ & 0.308 \\
   \hline 6 & $0.324\ Q^{-1/3}$ & $ 1.25\ Q^{1/3}$ & $0.1539\ Q^{-2/3}$ & 0.435 \\
   \hline 7 & $0.496 \ Q^{-1/4}$ & $ 1.14 \ Q^{1/4}$  & $ 0.245 \ Q^{-1/2}$ & 0.561 \\
   \hline $\cdots$ & $\cdots$ & $\cdots$ & $\cdots$ & $\cdots$ \\
   \hline $\infty$ & $ \varpropto \ Q^{-1/(D-3)}$ & $ \varpropto  \ Q^{1/(D-3)}$  & $ \varpropto  \ Q^{2/(D-3)}$ & $\cdots$ \\
  \hline
\end{tabular}
   \caption{The temperature, thermodynamical pressure, special volume and universal relation of critical points in diverse dimensions.}   \label{table:cp1}
 \end{table}
 We also study the phase transition in another perspective by plotting the $T - v$ transition fixing the thermodynamic pressure. The $T - v$ diagram is given by Fig.~\ref{vt56710}.
For fixing pressure, the critical points can also be obtained by
 \be
 \frac{\partial T}{\partial v} \Big|_{{\cal P}_c} = 0 \,, \quad  \frac{\partial^2 T}{\partial v^2} \Big|_{{\cal P}_c}  = 0 \,. \label{cp2}
 \ee
The results are the same as in Table~\ref{table:cp1}.

After obtaining the ${\cal P} -v$ and $T -v$ diagrams, together with analysis in the previous section, we can try to explain the ${\cal P} -v$ transition behaviors in EMD theory for different gauge coupling constants. In the cases $N=1, D=4,5,6,7$ dimensions, and $N=2$ in $D =4$ dimension which are coincident with gauge supergravities, the Van der Waals behaviors are absent because there is no extremal black hole in the range of $0 <N <N^\star$. In these cases, temperature cannot go to zero which is different from the standard Van der Waals behaviors in Fig.~\ref{vt56710}. In the cases $N=2$ in six and seven dimensions which are coincident with gauge supergravities, the phase transition behaviors can be viewed as Van der Waals-like behaviors. At high temperature, there are two branches of isotherms of which one is the same as the standard Van der Waals behavior and corresponds to positive pressure, and the other one is different and corresponds to negative pressure. As the temperature goes to zero, the negative branch disappears because the near-horizon geometries are also conformal AdS$_2 \times S^{D-2}$ in the range of $N^\star <N \le N^{RN}$. For cases $N =3, D=4$ and  $N =2, D=5$ which are the special cases of $N = N^\star$, as the threshold and the low bound of the existence of extremal black hole, have the same behaviors as the RN-AdS black holes.

\begin{figure}[h]
\centerline{
\ \ \ \ \ \includegraphics[width=0.45\linewidth]{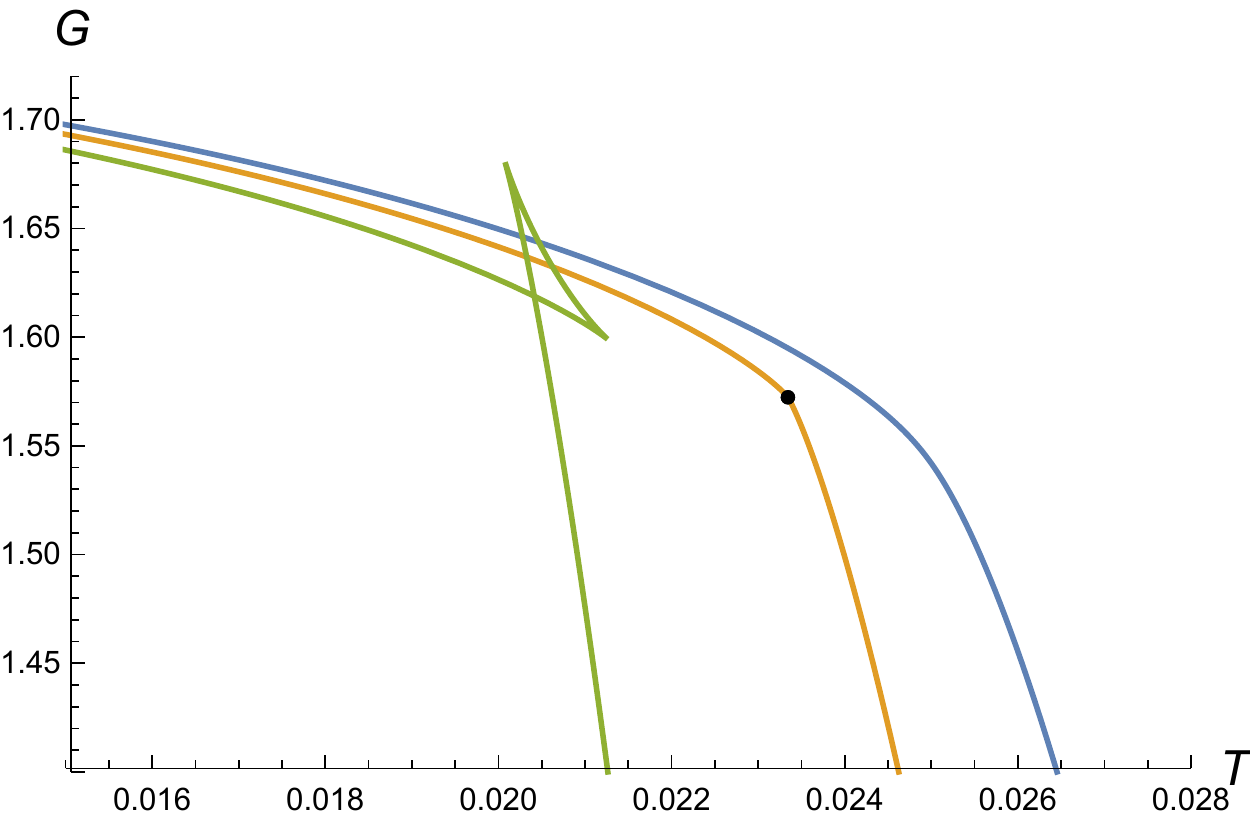}\ \ \ \
\includegraphics[width=0.45\linewidth]{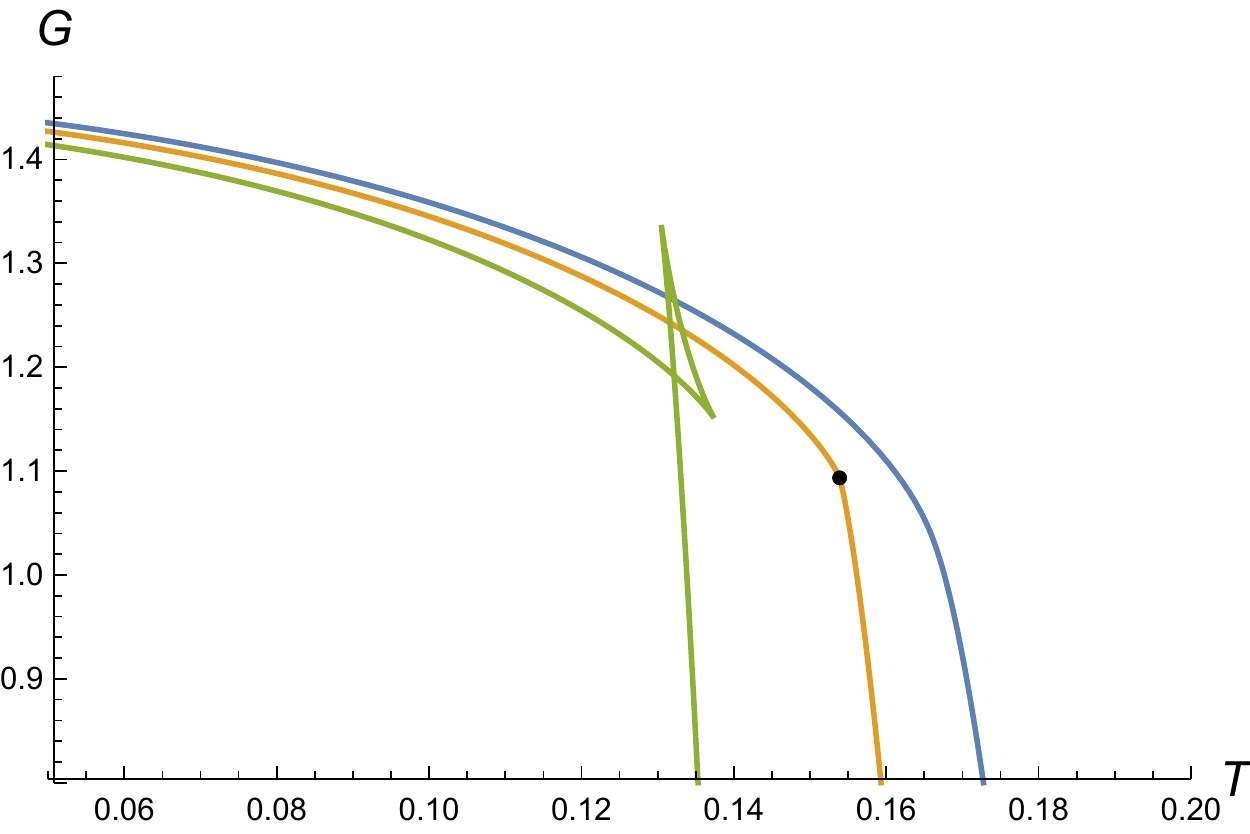}}
\centerline{
\ \ \ \ \ \includegraphics[width=0.45\linewidth]{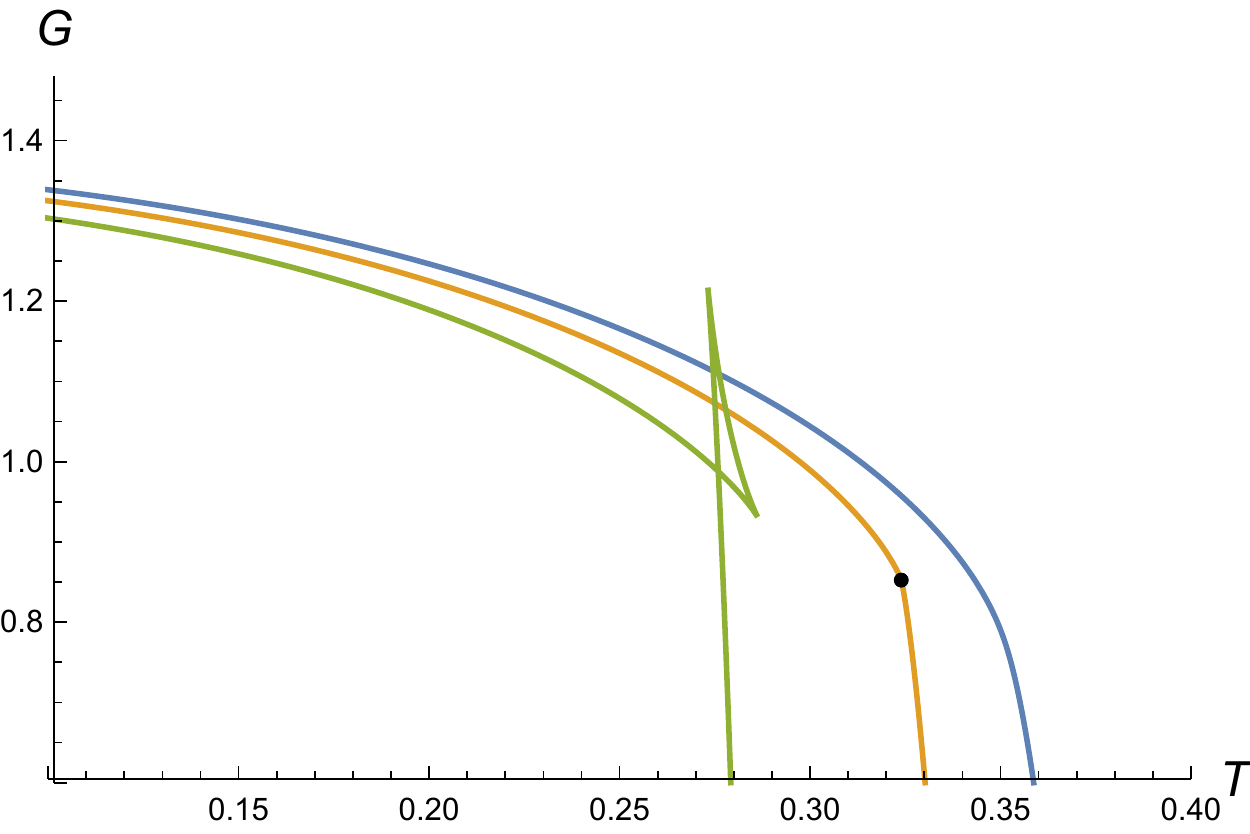}\ \ \ \
\includegraphics[width=0.45\linewidth]{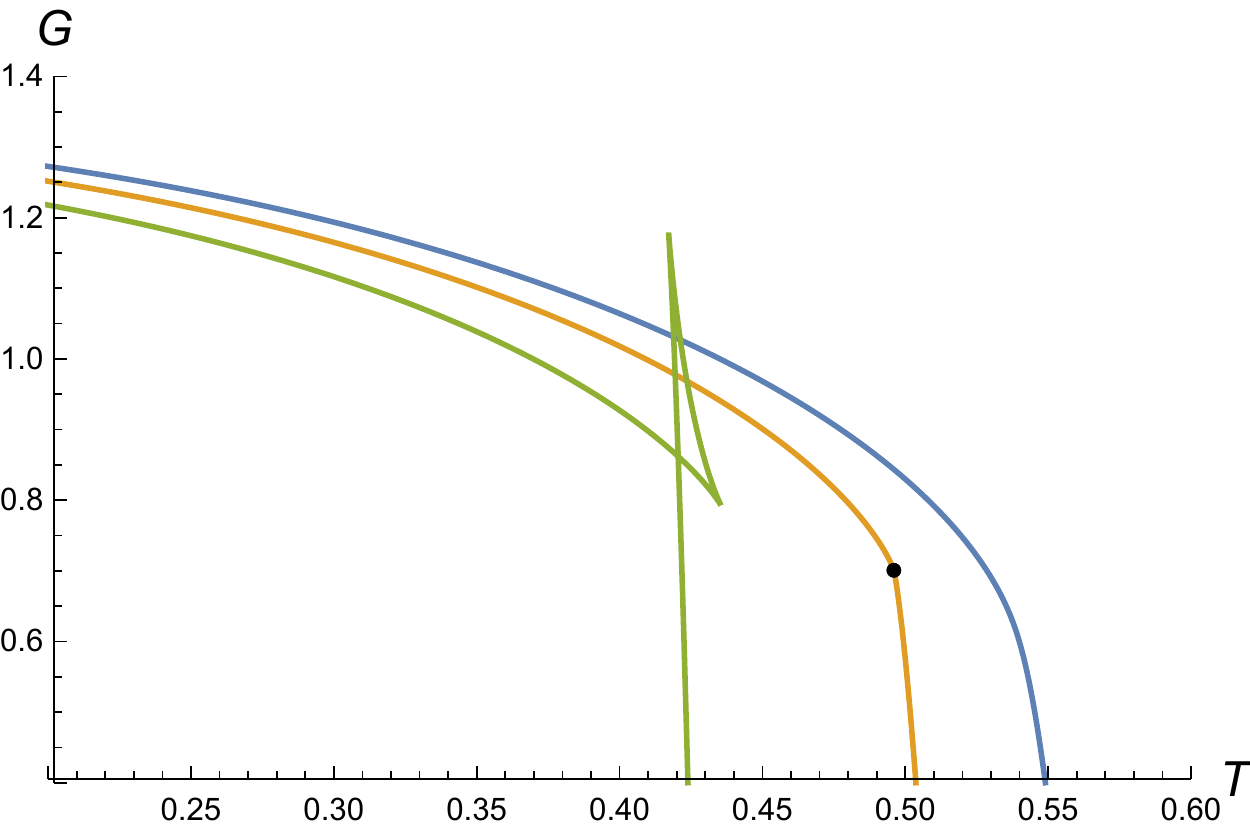}}
\caption{ Isobars in $G - T$ diagrams of charged dilatonic AdS black hole in diverse dimensions with $N=N^\star$. We set $Q=1$. Top left: $D=4$; top right: $D=5$; down left: $D=6$; down right: $D=7$.  The blue, orange and green lines represent isobars with ${\cal P} = 1.2\ {\cal P}_c, {\cal P}_c,$ and $ 0.7\ {\cal P}_c$ from top to bottom. The black points represent critical points.}
\label{gt56710}
\end{figure}

\begin{figure}[h]
\centerline{
\ \ \ \ \ \includegraphics[width=0.3\linewidth]{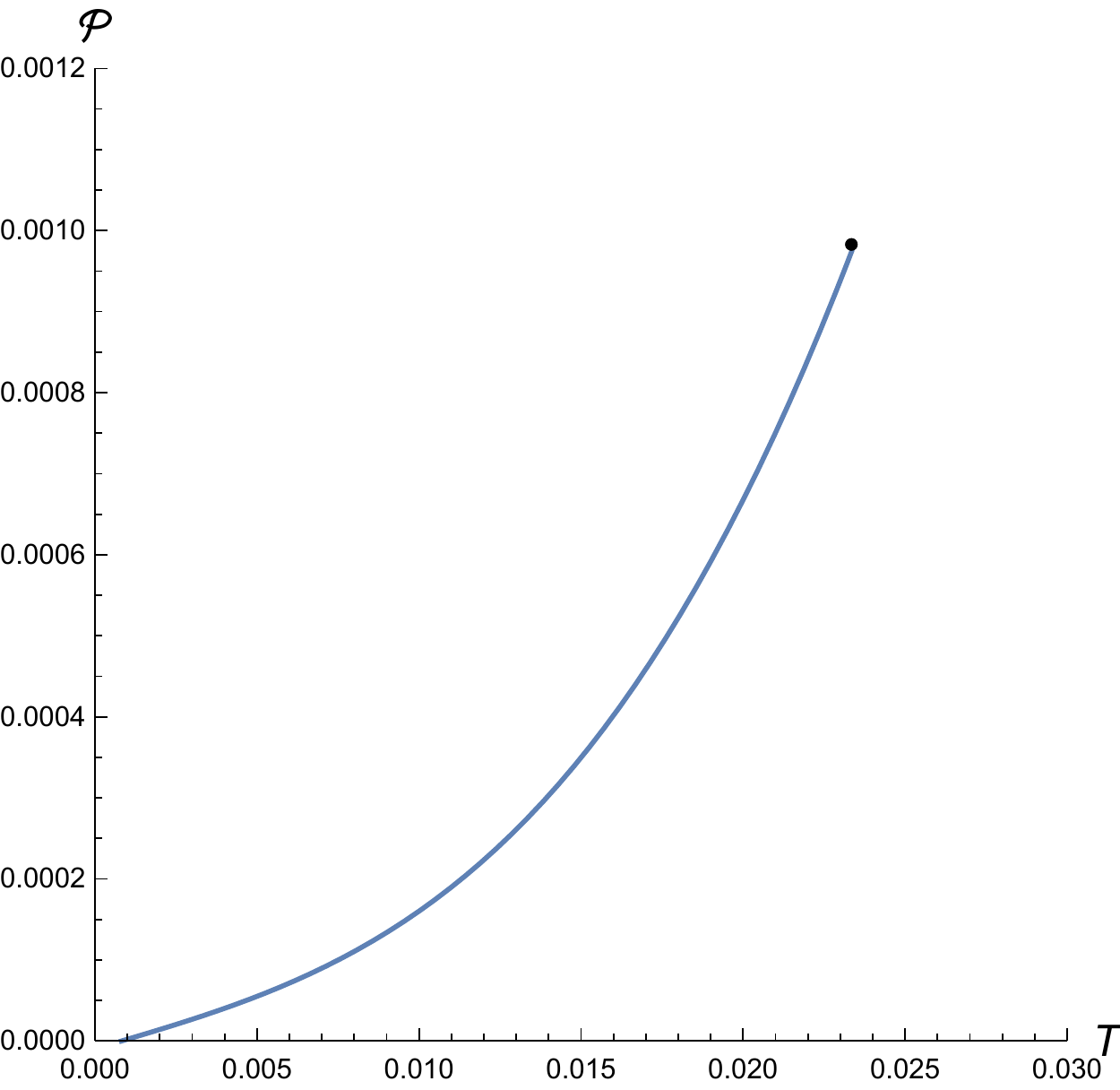}\ \ \ \
\includegraphics[width=0.3\linewidth]{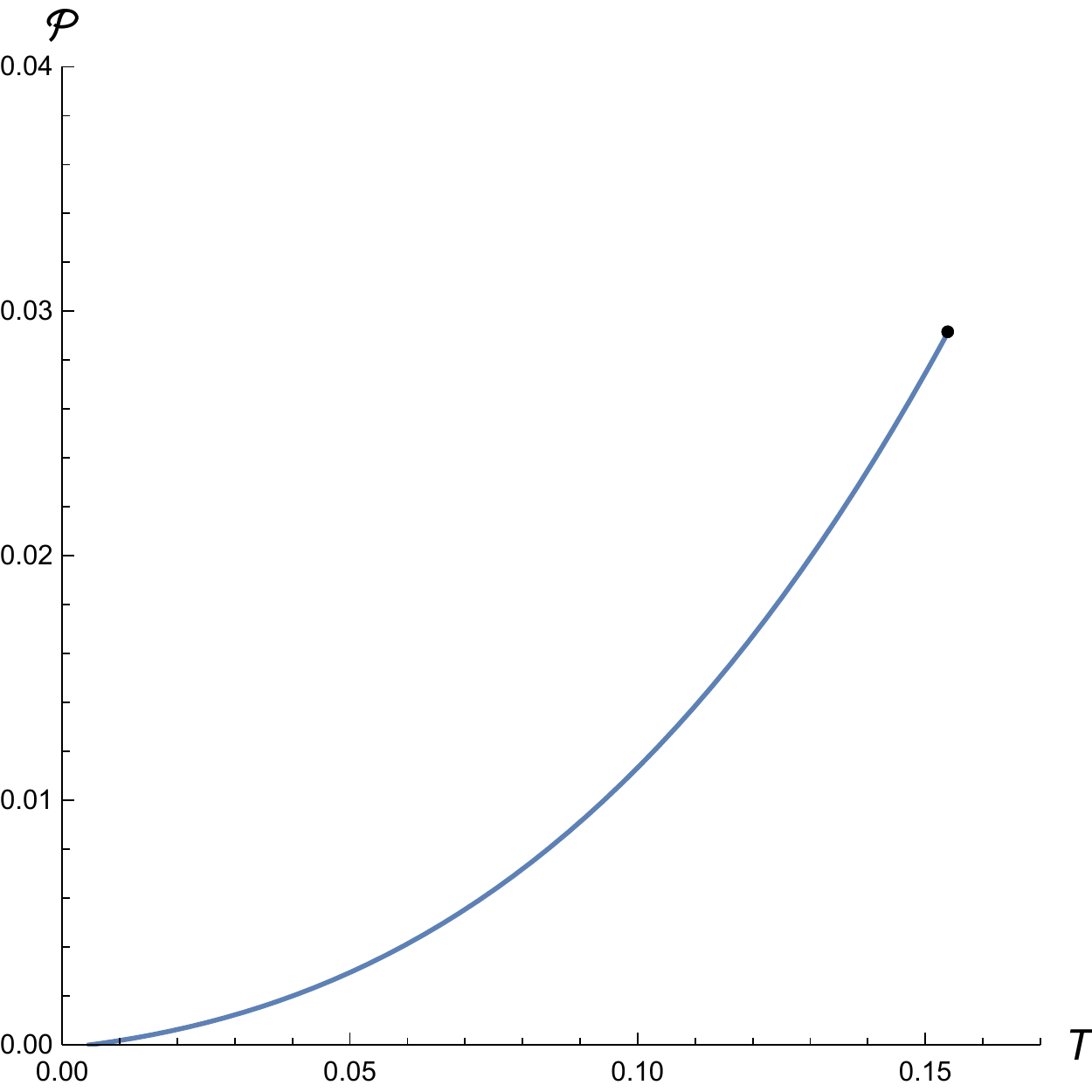}}
\centerline{
\ \ \ \ \ \includegraphics[width=0.3\linewidth]{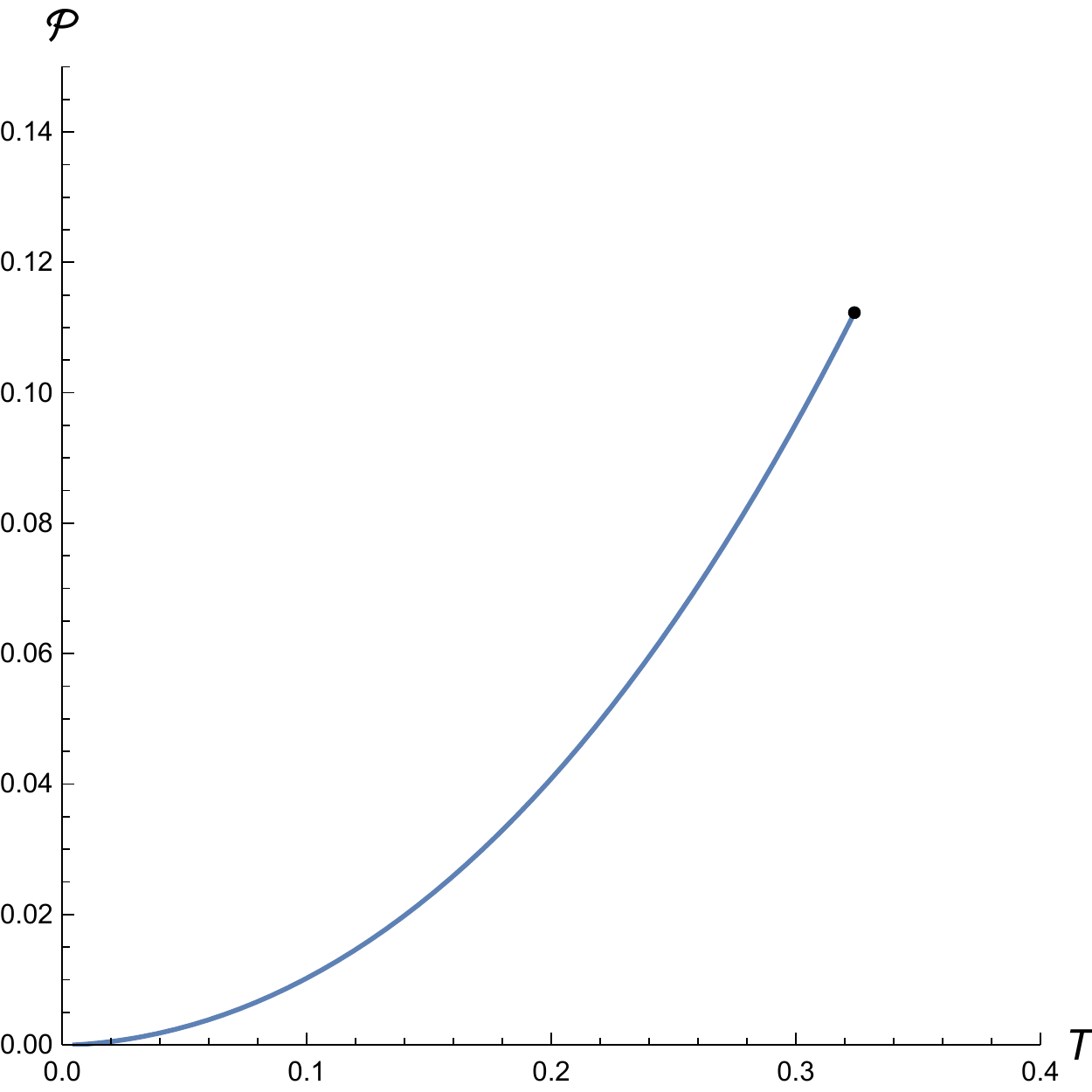}\ \ \ \
\includegraphics[width=0.3\linewidth]{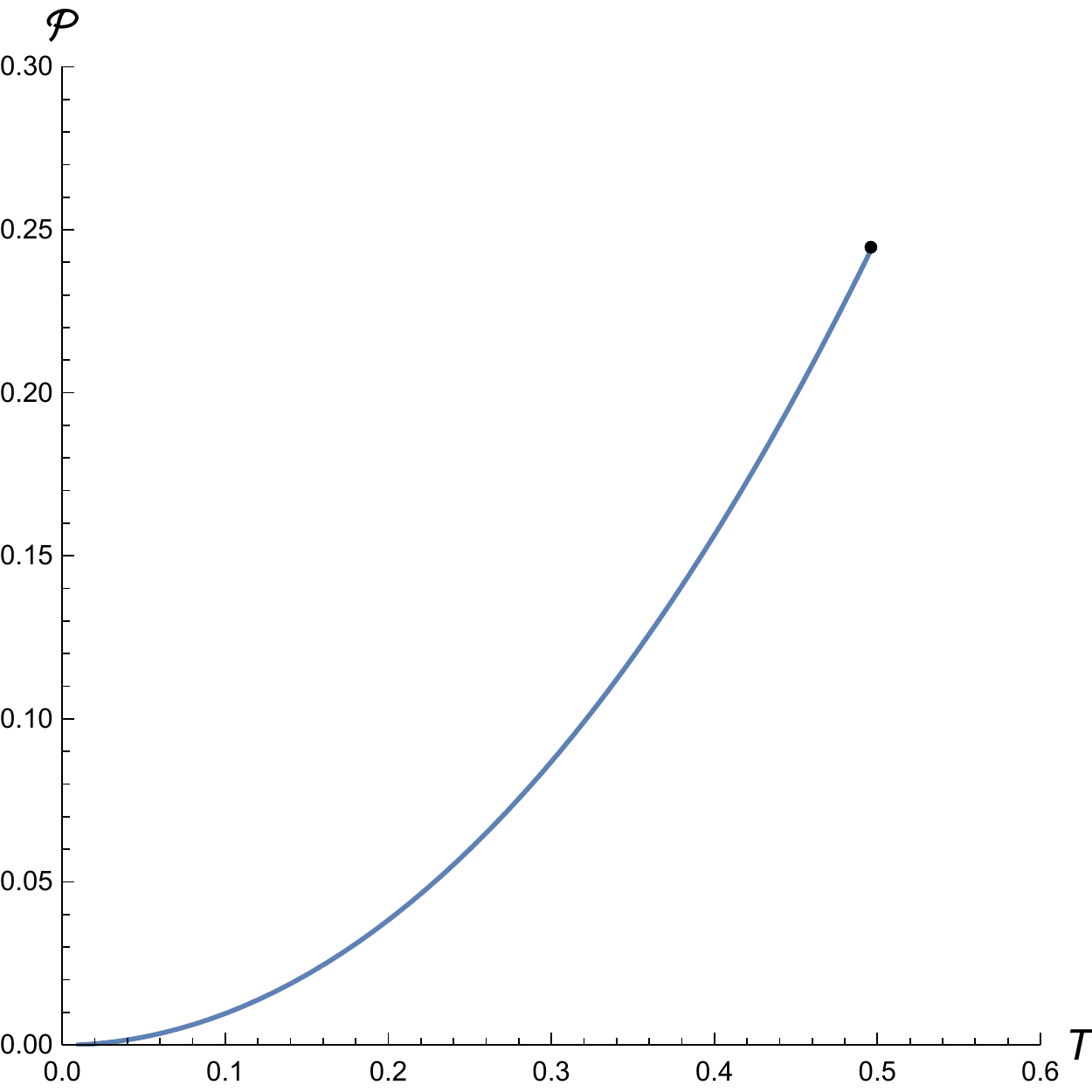}}
\caption{${\cal P} - T$ phase diagrams for diverse dimensions. Top left: $D=4$; top right: $D=5$; down left: $D=6$; down right: $D=7$. The black points are critical points at the end of the coexistence lines.}
\label{pt56710}
\end{figure}

In order to give the complete analysis of extended thermodynamics, we also study the Gibbs free energy, ${\cal P} -T$ plane and obtain critical exponents. The Gibbs free energy can be obtained simply by $G \equiv G(T, {\cal P })= M - T S$. According to~\cite{Kubiznak:2012wp}, we also plot the isobars of $G - T$ diagrams and ${\cal P} - T$ plane in Figs.~\ref{gt56710} and~\ref{pt56710}.
It can be viewed as small-big black hole transition similar to liquid-gas transition~\cite{Kubiznak:2012wp}. The critical exponents which characterizing the phase transition are given by
\be
\alpha =0 \,,\quad \beta = \frac{1}{2} \,,\quad \gamma =1 \,,\quad \delta = 3 \,.
\ee
which is the same as the standard Van der Waals phase transition~\cite{Kubiznak:2012wp}.

\subsection{$T^{-1} -S$ criticality }  \label{sec3.2}

\begin{figure}[h]
\centerline{
\ \ \ \ \ \includegraphics[width=0.45\linewidth]{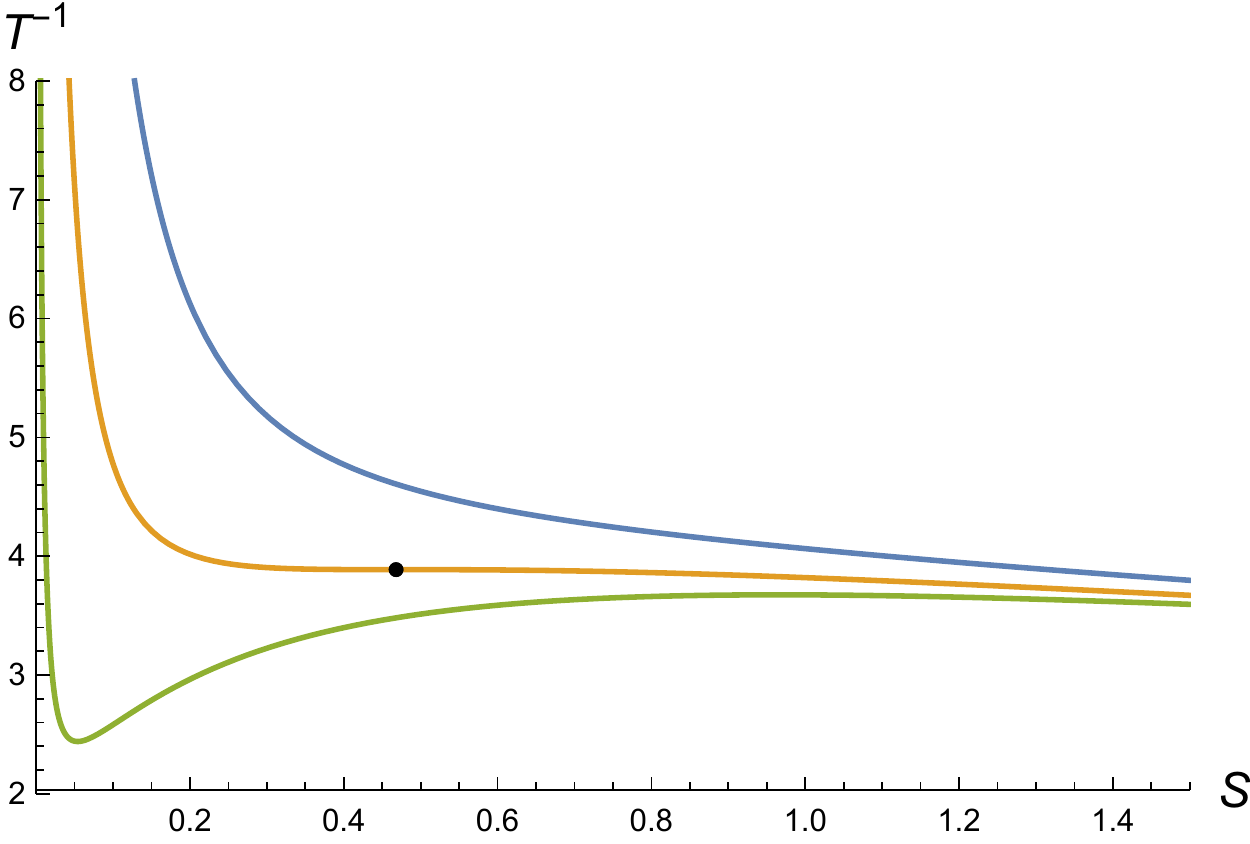}\ \ \ \
\includegraphics[width=0.45\linewidth]{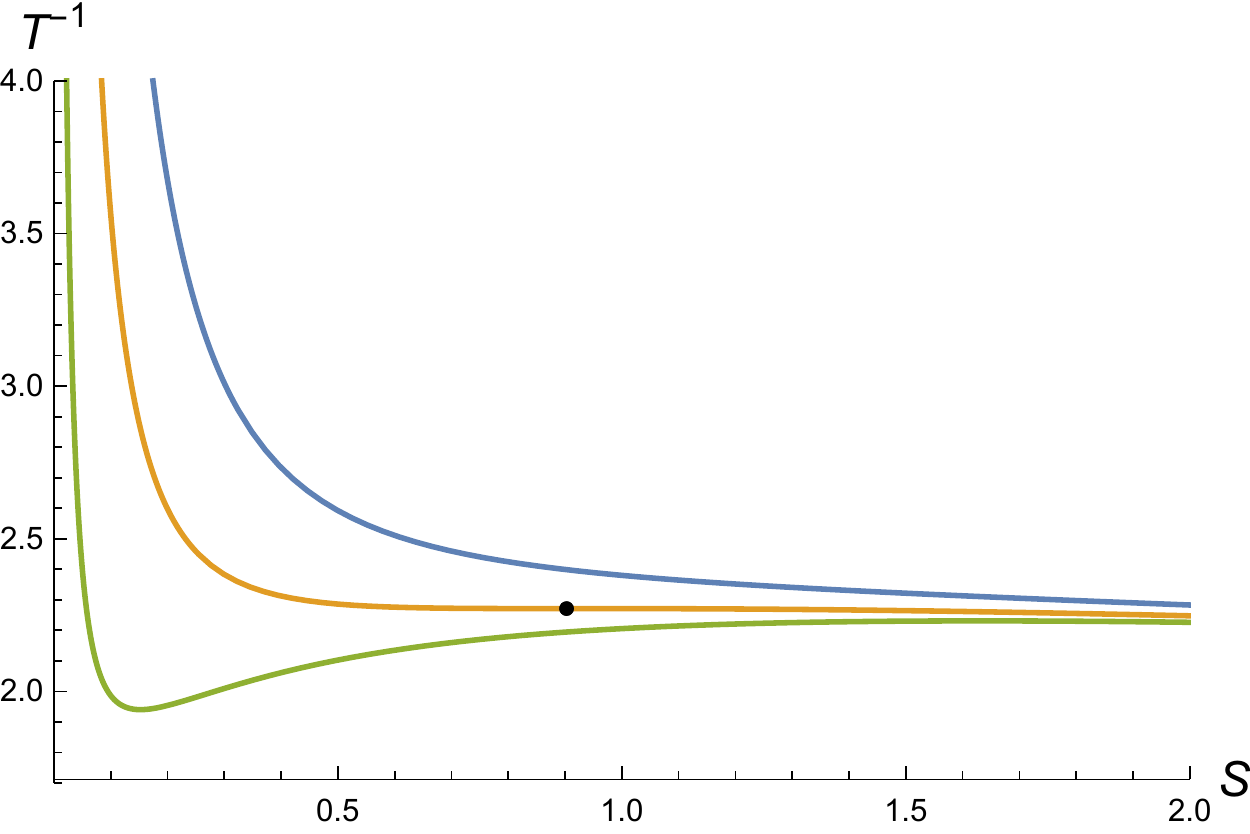}}
\centerline{
\ \ \ \ \ \includegraphics[width=0.45\linewidth]{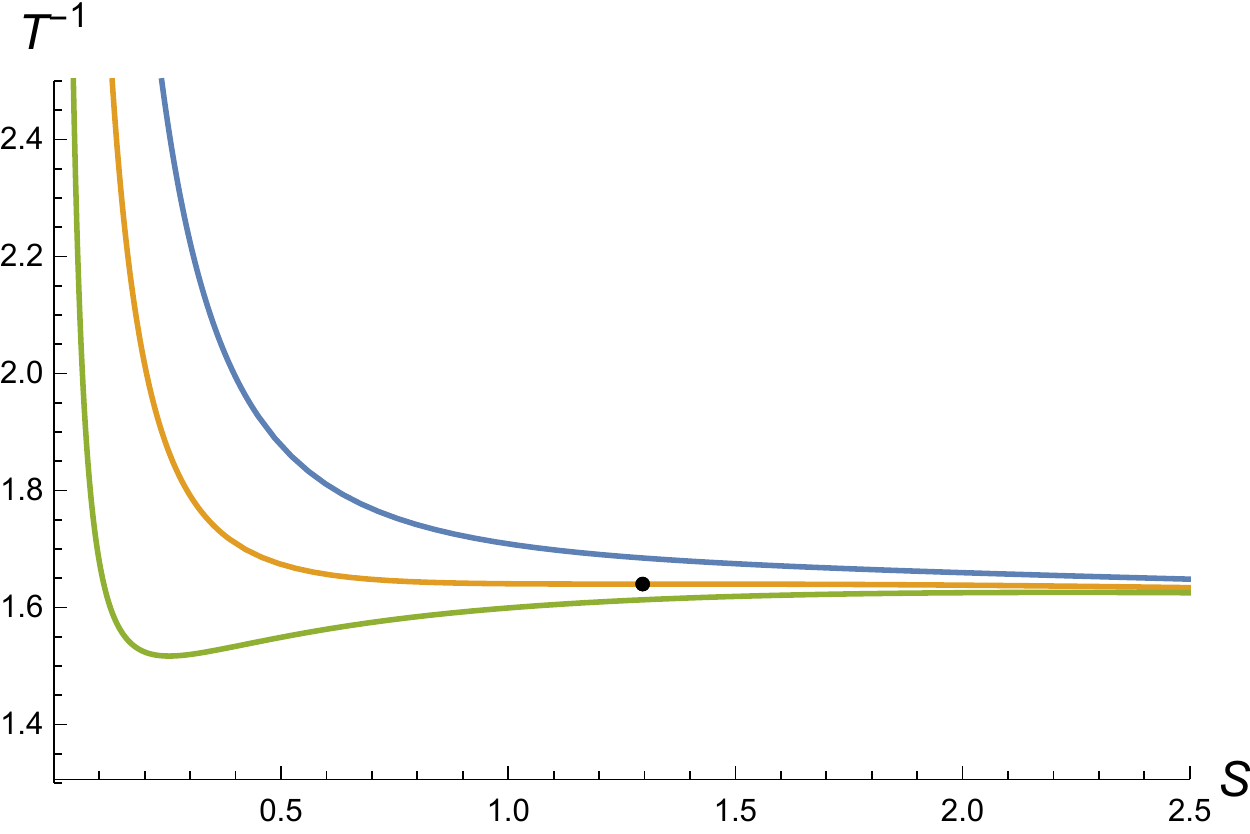}\ \ \ \
\includegraphics[width=0.45\linewidth]{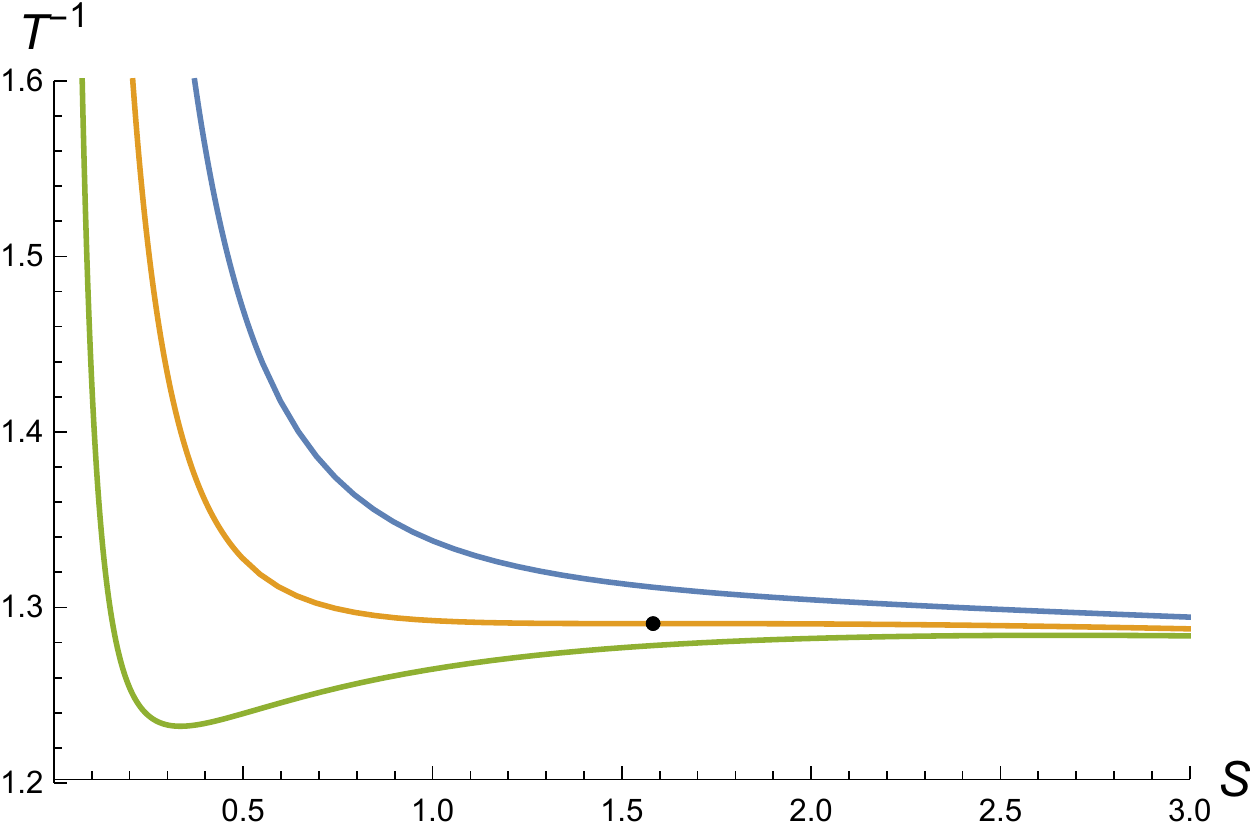}}
\caption{Iso-charges in $T^{-1} -S$ diagrams of charged dilatonic AdS black hole in diverse dimensions with $N=N^\star$. We set $g=1$. Top left: $D=4$; top right: $D=5$; down left: $D=6$; down right: $D=7$.  The blue, orange and green lines represent isobars with $Q = 1.5\ Q_c, Q_c,$ and $0.5\ Q_c$ from top to bottom. The black points represent critical points.}
\label{ggt56710}
\end{figure}

Now, we study the inverse temperature versus the Bekenstein-Hawking entropy transition by fixing the thermodynamical pressure independent with dimension, i.e. the gauge coupling constant $g$. From Ref.~\cite{Caceres:2015vsa}, the $T^{-1}-S$ transition of the STU black hole with three equal charges and the RN-AdS black hole have standard Van der Waals behaviors. Now we study the general dilatonic AdS black holes in EMD theory with the special dilaton coupling constant $N = N^\star$. The temperature, entropy and charge can be written as
\begin{align}
T &= \frac{D-3}{4 \pi } \Big[ (q+ R )^{-\frac{N^\star }{2}} R^{\frac{1}{2}} +N^\star g^2 (q + R)^{-\frac{D-5}{D-3}} R^{\frac{1}{2}}  \Big] \,, \label{temp3} \\
S &= \frac{\pi^{\frac{D-1}{2}} R^{\frac12}}{2 \, \Gamma(\frac{D-1}{2})} (q +R)^{\frac{N^\star }{2}} \,,  \label{pressure3} \\
Q &= \frac{(D-3) \pi^{\frac{D-3}{2}}}{8 \, \Gamma(\frac{D-1}{2})} \sqrt{ q N^\star (q +R +g^2 (q+R)^{N^\star})} \,,  \label{rho3}
\end{align}
It is also hard to obtain the exact EoS for $(T, S, Q)$ system. We study the phase transition by numeric method used before. To be specific, we set $g=1$. The $T^{-1} - S$ diagram is given by Fig.~\ref{ggt56710}.
 We also illustrate only $D = 4, 5, 6, 7$ dimensions to show $T^{-1} -S$ behaviors. And we find the higher dimensional cases have the same behaviors. From Fig.~\ref{ggt56710}, we find $T^{-1} -S$ transition have standard Van der Waals behaviors. As the entropy decreases to zero, the inverse temperature increases to infinity. As the entropy goes to infinity, the inverse temperature goes to zero. The corresponding critical point can be obtained by
 \be
 \frac{\partial { T}}{\partial S} \Big|_{Q_c} = 0 \,, \quad  \frac{\partial^2 { T}}{\partial S^2} \Big|_{Q_c}  = 0 \,. \label{cp3}
 \ee
The critical points are given in Table~\ref{table:cp2}
 \begin{table}[h]
 \centering
\begin{tabular}{|c|c|c|c|c|}
   \hline  $D$  & $Q_c$ & $T^{-1}_c$ & $S_c$ & $\frac{Q_{c} T^{-1}_{c}}{S_{c}}$ \\
   \hline 4 & $0.0907\ g^{-1}$ & $3.89\ g^{-1}$ & $0.468\ g^{-2}$ & 0.754    \\
   \hline 5 & $0.122\ g^{-2}$ & $0.227\ g^{-1}$ & $0.903\ g^{-3}$ & 0.307 \\
   \hline 6 & $0.150\ g^{-3}$ & $1.64\ g^{-1}$ & $1.30\ g^{-4}$ & 0.190 \\
   \hline 7 & $0.168\ g^{-4}$ & $ 1.29\ g^{-1}$ & $1.58\ g^{-5}$ & 0.137 \\
   \hline $\cdots$ & $\cdots$ & $\cdots$ & $\cdots$ & $\cdots$ \\
   \hline $\infty$ & $ \varpropto \ g^{-(D-3)}$ & $ \varpropto  \ g^{-1}$  & $ \varpropto  \ g^{-(D-2)}$ & $\cdots$ \\
  \hline
\end{tabular}
   \caption{The charge, temperature, Bekenstein-Hawking entropy and universal relation of critical points in diverse dimensions.}   \label{table:cp2}
 \end{table}
We can also obtain the corresponding critical exponent $ \delta_1$ characterizing the transition from
\be
T^{-1}- T_c^{-1} \varpropto |S - S_c|^{\delta_1} \quad \Rightarrow \quad \delta_1 = 3 \,,
\ee
which is the same as the critical exponent $\delta$ in fixing the charge ensemble by the numerical method.

\section{HEE }  \label{sec4}

After studying the extended thermodynamics of charged dilatonic AdS black hole systems with near-horizon geometry conformal to AdS$_2 \times S^{D-2}$ in the bulk view, we find both ${\cal P} -v$ and $T^{-1} - S$ transitions have the standard Van der Waals behaviors.  Now we consider whether HEE undergoes the same transition behaviors in the dual field theory. We start with a review of Ryu-Takayanagi (RT) prescription following e.g.~\cite{Ryu:2006bv, Rangamani:2016dms} and calculate the HEE of charged dilatonic AdS black holes in general EMD theory with string-inspired potential. Then we plot the transition behaviors.

In the frame of field theory, for a bipartitioning of Hilbert space ${\cal H}_\alpha$ system, it can be decomposed  into two separate tensor factors
\be
\otimes_\alpha {\cal H}_\alpha \cong {\cal H}_{\cal A} \otimes {\cal H}_{{\cal A}^c} \,,
\ee
where ${\cal A}$ and ${\cal A}^c$ are the region within  and  outside the boundary, entangling surface, $\partial {\cal A}$. We can construct an operator, reduced density matrix, $\rho_{\cal A}$ that acts on ${\cal H}_{\cal A}$ by tracing out the other ${\cal H}_{{\cal A}^c}$,
\be
\rho_{\cal A} = \textup{Tr}_{{\cal A}^c} (\Ket{\Psi}\Bra{\Psi}) \,,
\ee
where the pure quantum state $\Ket{\Psi}$ is an element of tensor product Hilbert space. In order to quantifying the amount of entanglement existing in the state $\Ket{\Psi}$ by spatial decomposition. We can introduce the von Neumann entropy of reduced density matrix,
\be
S = -\textup{Tr}_{\cal A} (\rho_{\cal A} \log \rho_{\cal A}) \,,
\ee
which is also referred to as entanglement entropy (EE).

Now we consider the EE in AdS/CFT correspondence frame. Consider a CFT living on boundary ${\cal B}$ whose geometry is described by metric $h_{\mu\nu}$. We can take a region {\cal A} with entangling surface $\partial {\cal A}$ to lie on some Cauchy slice which is the subset of boundary, so  EE can be computed holographically at some constant time slice. We can choose a coordinate systems $\xi^a$ on the surface $\partial {\cal A}$ via a set of mappings $x^\mu(\xi^a)$. The induced metric is given by
\be
ds^2_{\partial {\cal A}} = h_{\mu\nu} dx^\mu dx^\nu = h_{\mu \nu} \frac{\partial x^\mu}{\partial \xi^a} \frac{\partial x^\nu}{\partial \xi^b} d\xi^a d\xi^b \,.
\ee
Assuming there is a bulk $\cal M$ with metric $g_{\mu\nu}$ dual to the boundary CFT, and according to RT prescription, we need to find a codimension-two extremal surface ${\cal E}_{\cal A}$ satisfying boundary condition ${\cal E}_{\cal A}|_{\cal B} = \partial {\cal A}$ in bulk spacetime. The induced metric with extremal area is given by
\be
ds^2_{{\cal E_A}} = g_{\mu\nu} dx^\mu dx^\nu  = g_{\mu \nu} \frac{\partial x^\mu}{\partial \xi^a} \frac{\partial x^\nu}{\partial \xi^b} d\xi^a d\xi^b \,.
\ee
So we have
\be
h_{\mu \nu} = g_{\mu \nu} \frac{\partial x^\mu}{\partial \xi^a} \frac{\partial x^\nu}{\partial \xi^b} \,.
\ee
We should find the one of the extremal surfaces satisfying homology requirement that has the smallest area. The HEE is given by the area of this minimal extreme surface in some sense similar to Bekenstein-Hawking entropy formula,
\be
S_{\cal A} = \frac{\textup{Area}({\cal E}_{\cal A})}{4\ G^D} \,,
\ee
where $G^D$ is the $ D$ dimensional Newton constant and we set $4\ G^D =1$ in reminder discussion.

\begin{figure}[h]
\centerline{
\ \ \ \ \ \includegraphics[width=0.45\linewidth]{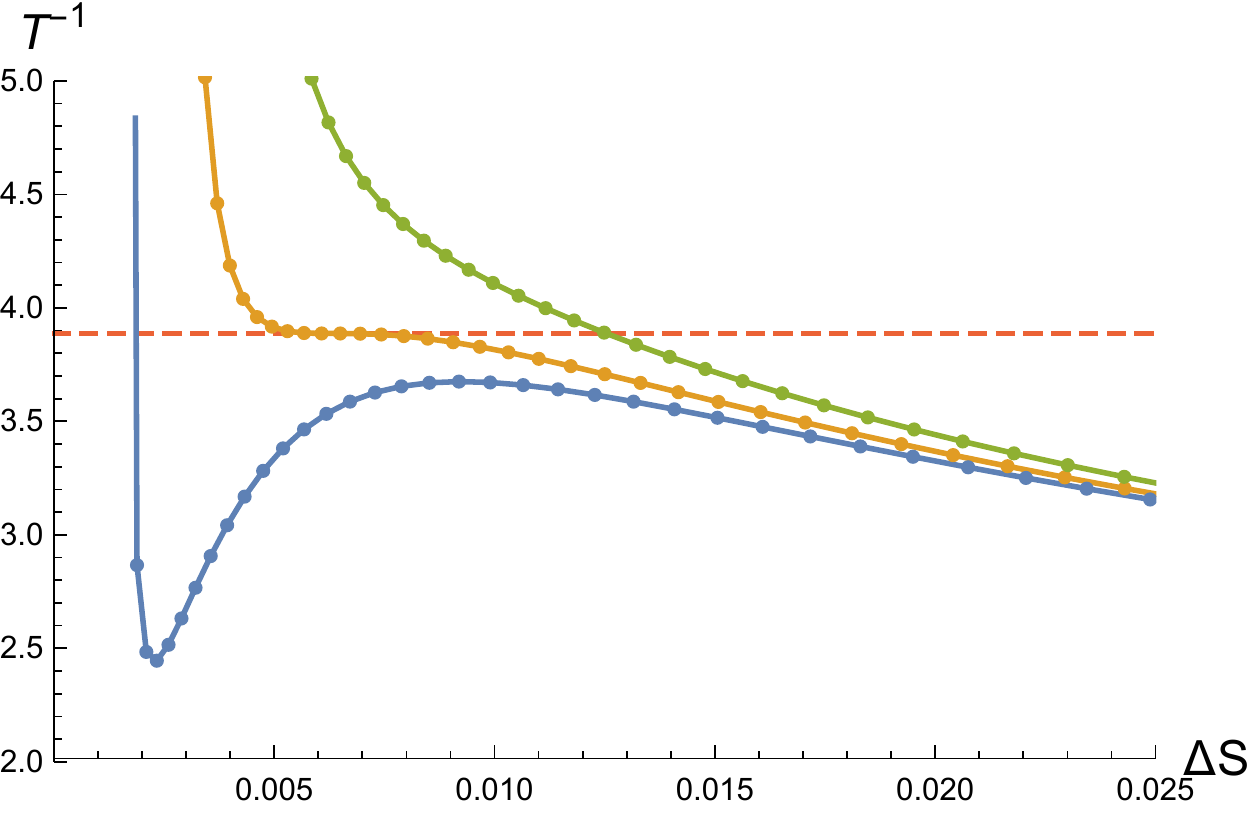}\ \ \ \
\includegraphics[width=0.45\linewidth]{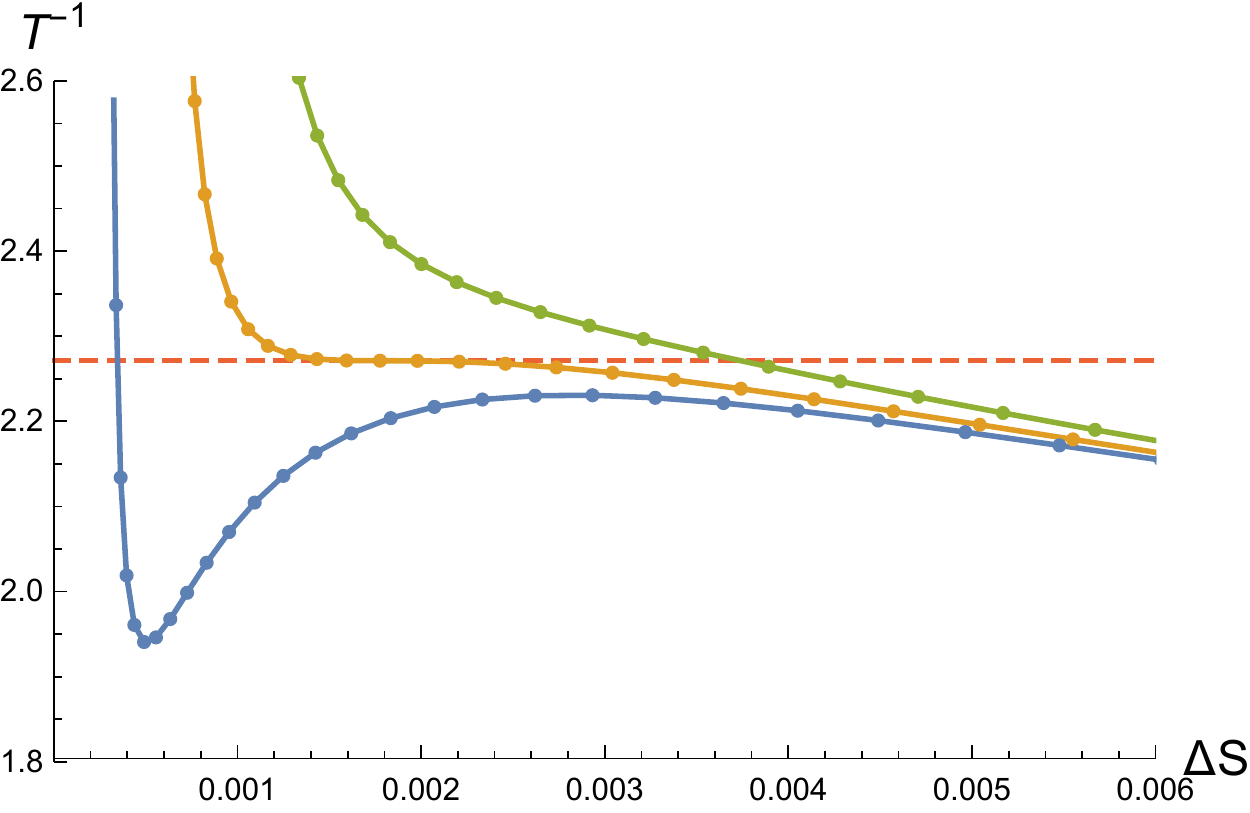}}
\centerline{
\ \ \ \ \ \includegraphics[width=0.45\linewidth]{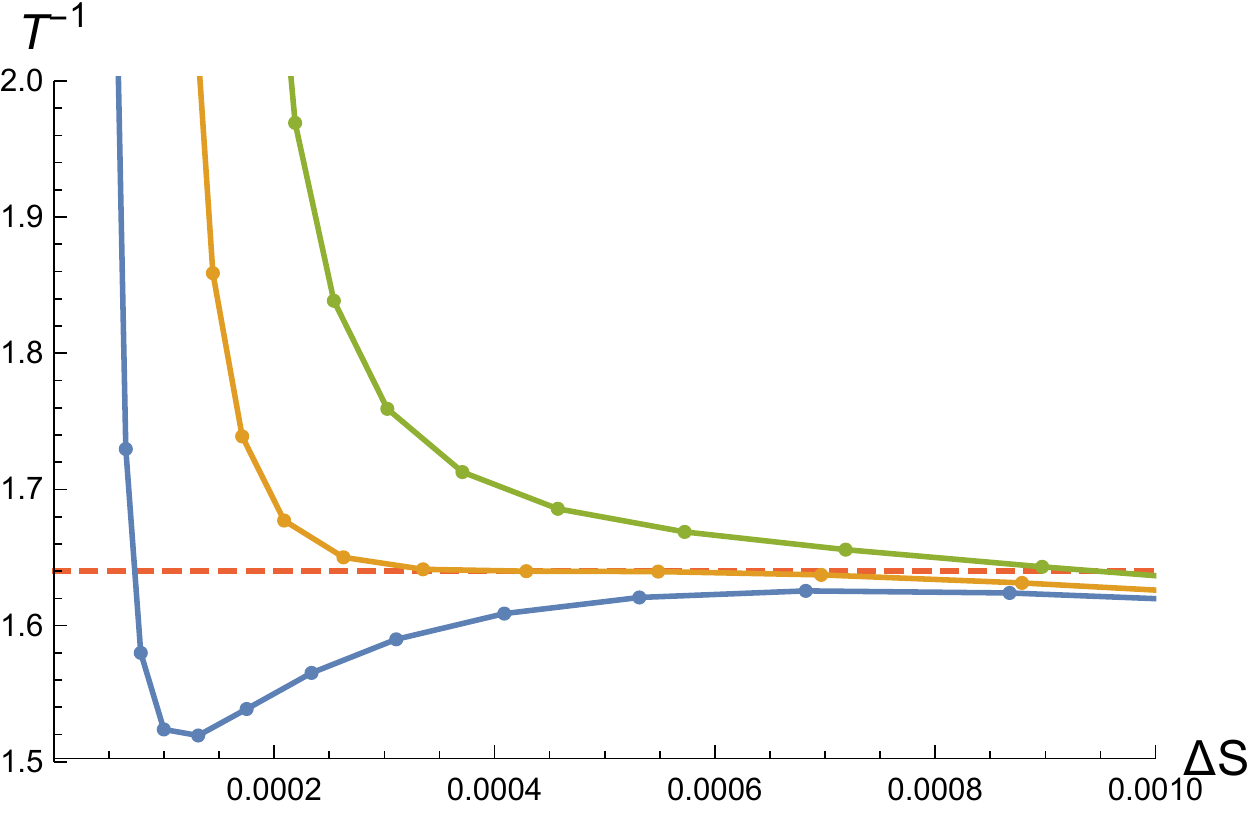}\ \ \ \
\includegraphics[width=0.45\linewidth]{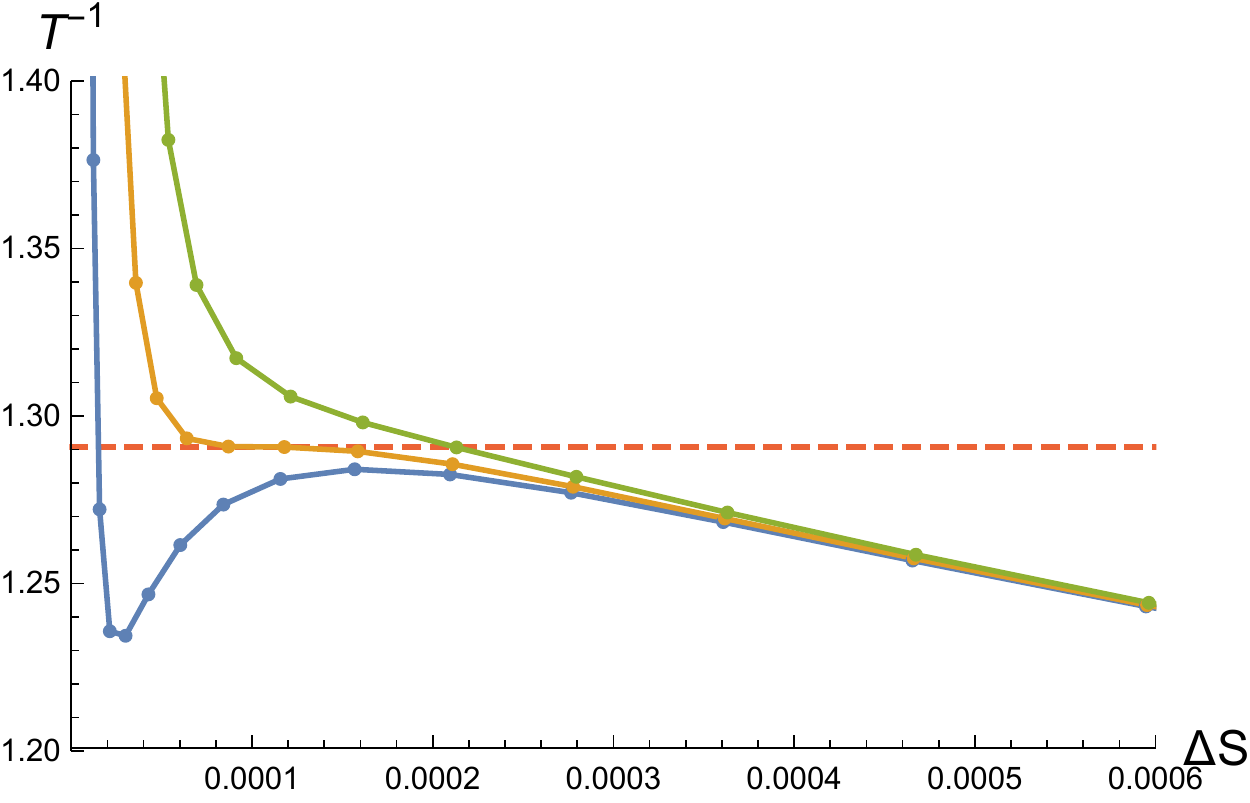}}
\caption{Iso-charges in $T^{-1} -\delta S$ diagrams of charged dilatonic AdS black holes in diverse dimensions with $N=N^\star$. We set $g=1$. Top left: $D=4$; top right: $D=5$; down left: $D=6$; down right: $D=7$.  The green, orange and blue lines represent isobars with $Q = 1.5\ Q_c, Q_c,$ and $0.5\ Q_c$ from top to bottom. The dash lines represent inverse critical temperatures.}
\label{hee56710}
\end{figure}

Consider a constant time slice and take a spherical region within and outside the entangling surface as $\theta = \theta_0$. We must find the extremal area surface in the bulk, the coordinates $\xi^a = (\theta, x_1, \dots, x_{D-3})$. Because the geometry of bulk region is spherical symmetric, so the radial coordinate of the extremal surface in bulk region is only dependent on $\theta$. We assume the extremal surface running from $r=r_0$ at $\theta=0$ to $r \rightarrow \infty$ at $\theta = \theta_0$. Due to the extremal area has the minimal area, we can treat the entropy as an action
\be
S_{\cal A} = \Omega_{D-3} \int_{0}^{\theta_0} d\theta \sin^{D-3}(\theta) \ r^{D-3} h^{\frac{N^\star}{2}} \sqrt{r^2 + \frac{\dot{r}^2}{f}} \,,
\ee
where $\Omega_{D-3} = 2 \pi^{(D-2)/2}/\Gamma((D-2)/2)$, and extremize it through Euler-Lagrangian equation with respect to $r(\theta)$. In order to avoid the divergence, we can integrate to some cutoff $\theta_c \lesssim \theta_0$. And in order to filter out the thermal entropy, we choose the entangling surface as a small value for $\theta_0$. To be specific, we choose $\theta_0 = 0.2 $, and the cutoff $\theta_c = 0.199 $. Since the EE is UV-divergent, it needs to be regularized. The minimal area of AdS vacuum can exactly be given by
\be
r = \frac{1}{g} \left( \frac{\cos^2\theta}{\cos^2\theta_0} -1 \right)^{-\frac12}  \,.
\ee
The minimal area of black hole can be calculated numerically. Then we subtract the entropy of extremal surface in AdS vacuum from the black hole. The result $\Delta S$ is so called renormalized entanglement entropy.
The $T^{-1} - \Delta S$ diagram is given by Fig.~\ref{hee56710}.
From Fig.~\ref{hee56710}, it can be seen that HEE undergoes the same Van der Waals transition behavior too. The above transition behavior would not change as the choice of the value of $\theta_0$ since we check it for different values. The critical exponent for the above transition behavior can be calculated numerically by
\be
T^{-1}- T_c^{-1} \varpropto |\Delta S - \Delta S_c|^{\delta_2} \quad \Rightarrow \quad \delta_2 = 3 \,,
\ee
which is the  same as the critical exponent $\delta$ of the ${\cal P} - v$ transition.

\section{Conclusions} \label{sec5}

As the first step in studying the holographic properties of the Van der Waals transition of charged dilatonic AdS black holes, we study the extended thermodynamics in EMD theory with string-inspired potential in general dimensions. By studying both the ${\cal P}-v$ transition in fixing the charge ensemble and the $T^{-1}-S$ transition in fixing the pressure ensemble in extended phase space, we find there exist standard Van der Waals transitions for a special class of black holes with near-horizon geometry conformal to AdS$_2 \times S^{D-2}$ with special dilaton coupling constant $N = N^\star$ which is different from its counterparts with other dilaton coupling constants. Instead of Bekenstein-Hawking entropy, we find HEE also undergoes the same transition behavior. The near-horizon geometries of  RN-AdS black holes are AdS$_2 \times S^{D-2}$, and the near-horizon geometries of the special class of charged dilatonic AdS black holes with dilaton coupling constant $N = N^\star$ are conformal to AdS$_2 \times S^{D-2}$ in the extremal or nearly extremal limit. Our results demonstrate that the near-horizon geometry AdS$_2 \times S^{D-2}$ is not necessary for the existence of the Van der Waals transition. The near-horizon conformal to AdS$_2 \times S^{D-2}$ is enough to realize the standard Van der Waals behaviors.

There are many further considerations in our work. First, based on the conjecture of the holographic interpretation of the Van der Waals behaviors~\cite{Johnson:2014yja}, it would be interesting to study the holographic heat engines of this special class of black holes in EMD theory. Second, though we find the black hole with similar near-horizon geometry has the same transition behaviors,  we cannot say definitively that  the similar near-horizon geometry is the real reason for the existence of Van der Waals behaviors. The thermodynamic quantities are coupled with each other and we cannot obtain the EoS exactly. So it is worth it to study the deep reason and check more black holes such as rotating~\cite{Wu:2011zzh}, dyonic~\cite{Lu:2013ura, Chow:2013gba, Wu:2015ska}, and accelerating~\cite{Lu:2014sza} black holes and so on, in EMD theory and $\omega$-deformed gauged KK supergravity. Because the dilatonic black hole can be lifted in M-branes and D$p$-branes~\cite{Tseytlin:1996bh}, we can also study the extended thermodynamics in the M-brane background~\cite{Chabab:2015ytz}.  Third, because we obtain the HEE of the special class dilatonic charged AdS black holes, it is also worth it to study some other entanglement quantities, such as  Wilson loops and correlation functions.

\section*{Acknowledgement }

S.L.L is grateful to Xiao-Xiong Zeng, Hua-Kai Deng and Hong-Da Lyu for kind discussions. He thanks especially H. L\"u for many suggestions and proofreading. This work is supported in part by NSFC grants No.~11575022, No.~11175016, No.~11475024, and No.~11875200 and the Graduate Technological Innovation Project of Beijing Institute of Technology.

\end{document}